\documentclass[graybox]{svmult}

\usepackage{mathptmx}       
\usepackage{helvet}         
\usepackage{courier}        
\usepackage{type1cm}        
\usepackage{makeidx}         
\usepackage{graphicx}        
\usepackage{multicol}        
\usepackage[bottom]{footmisc}

\makeindex             

\usepackage{amsmath,amssymb}
\usepackage{mathtools}
\usepackage{hhline}

\usepackage{amsthm}
\theoremstyle{plain}

\usepackage{calrsfs}
\DeclareMathAlphabet{\pazocal}{OMS}{zplm}{m}{n}
\renewcommand{\mathcal}[1]{\pazocal{#1}}

\usepackage{accents}
\newcommand{\ubar}[1]{\underaccent{\bar}{#1}} 

\DeclareSymbolFont{bbold}{U}{bbold}{m}{n}
\DeclareSymbolFontAlphabet{\mathbbold}{bbold}

\newcommand{\onev}{\mathbbold{1}}
\newcommand{\abs}[1]{\lvert #1 \rvert}
\newcommand{\dN}{\,dN}

\DeclareMathOperator*{\minimize}{\text{\upshape minimize}}
\DeclareMathOperator*{\subjectto}{\text{subject to}}
\DeclareMathOperator{\sgn}{sgn}
\DeclareMathOperator*{\col}{col}

\newcommand{\resourceGraphWidth}{.405\textwidth}
\newcommand{\betweenResourceGraphWidth}{.05\linewidth}

\begin{document}

\title*{Optimal Containment of Epidemics in Temporal and Adaptive Networks}
\titlerunning{Containing Epidemic Outbreaks in Temporal and Adaptive Networks} 
\author{Masaki Ogura and Victor M.~Preciado}
\institute{University of Pennsylvania, 3330 Walnut Street, Philadelphia, Pennsylvania 19104, USA, \email{\{ogura, preciado\}@seas.upenn.edu}
}
%
%
\maketitle


\abstract*{In this chapter, we overview a recent progress toward the containment
of disease spread taking place in temporal and adaptive networks. We
specifically focus on the optimal resource distribution problems to eradicate
epidemic outbreaks over the following three models of temporal and adaptive
networks: Markovian temporal networks, aggregated-Markovian edge-independent
temporal networks, and Adaptive SIS models. For each model, we present rigorous
and tractable frameworks based on convex optimizations for determining efficient
manners of distributing resources over networks to maximally accelerate the
extinction of disease spreads. The effectiveness of the frameworks are
illustrated by numerical simulations based on the temporal versions of Zachary
Karate Club Network.}

\abstract{In this chapter, we focus on the problem of containing the spread of diseases taking place in both \emph{temporal} and \emph{adaptive} networks (i.e., networks whose structure `adapts' to the state of the disease). We specifically focus on the problem of finding the optimal allocation of containment resources (e.g., vaccines, medical personnel, traffic control resources, etc.) to eradicate epidemic outbreaks over the following three models of temporal and adaptive networks: (i) Markovian temporal networks, (ii) aggregated-Markovian temporal networks, and (iii) stochastically adaptive models. For each model, we present a rigorous and tractable mathematical framework to efficiently find the optimal distribution of control resources to eliminate the disease. In contrast with other existing results, our results are not based on heuristic control strategies, but on a disciplined analysis using tools from dynamical systems and convex optimization.}

\section{Introduction}

The containment of spreading processes taking place in complex networks is a
major research area with applications in social, biological, and technological
systems~\cite{Watts1998,Newman2006,BBV08}. The spread of information in on-line
social networks, the evolution of epidemic outbreaks in human contact
networks, and the dynamics of cascading failures in the electrical grid are
relevant examples of these processes. While major advances have been made in
this field (see, for example, \cite{Nowzari2015a,Pastor-Satorras2015a} and references therein), most current results are specifically tailored to study spreading processes taking place in \emph{static}
networks. Cohen et al. \cite{CHB03} proposed
a heuristic vaccination strategy called \emph{acquaintance immunization
policy} and proved it to be much more efficient than random vaccine
allocation. In \cite{BCGS10}, Borgs et al. studied theoretical
limits in the control of spreads in undirected network with a non-homogeneous
distribution of antidotes. Chung et al. \cite{CHT09}
studied a heuristic immunization strategy based on the PageRank vector
of the contact graph.
Preciado et al. \cite{PZEJP13,Preciado2014} studied the problem of determining the optimal
allocation of control resources over static networks to efficiently eradicate
epidemic outbreaks described by the networked SIS model. This work was later
extended in~\cite{PZ13,PSS13,PJ09,NPP14,XP14,NOPP15,WNPP15,Watkins2015} by considering more general epidemic models.
Wan et al. developed in \cite{WRS08} a control theoretic framework for disease spreading, which has been recently extended to the case of sparse control strategies in \cite{TRW15}.
Optimal control problems over networks have also been considered in \cite{KSA11} and \cite{KB14}.
Drakopoulos et al. proposed in \cite{DOT14} an efficient curing policy based on graph cuts.
Decentralized algorithms for epidemic control have been proposed in \cite{RM15} and
in~\cite{Trajanovski2015a} using a game-theoretic framework to evaluate the effectiveness of protection
strategies against SIS virus spreads. An optimization framework to achieve 
resource allocations that are robust to stochastic uncertainties in nodal
activities was proposed in~\cite{Ogura2015g}.

Most epidemic processes of practical interest take
place in \emph{temporal networks}~\cite{Masuda2013}, having time-varying
topologies~\cite{Holme2015b}. In the context of temporal networks,
we are interested in the interplay between the epidemiological dynamics on
networks (i.e., the dynamics of epidemic processes taking place in the network)
and the dynamics of networks (i.e., the temporal evolution of the network
structure). Although the dynamics on and of networks are usually studied
separately, there are many cases in which the evolution of the network structure
is heavily influenced by the dynamics of epidemic processes taking place in the
network. This can be illustrated by a phenomenon called social
distancing~\cite{Bell2006,Funk2010}, where healthy individuals avoid contact
with infected individuals in order to protect themselves against the disease. As
a consequence of social distancing, the structure of the network adapts to the
dynamics of the epidemics taking place in the network. Similar adaptation
mechanisms have been studied in the context of the power grid~\cite{Scire2005},
biological systems~\cite{Schaper2003} and on-line social
networks~\cite{Antoniades2013}.

We can find a plethora of studies dedicated to the \emph{analysis} of epidemic spreading processes
over temporal networks based on either extensive numerical
simulations~\cite{Vazquez2007,Karsai2011,Holme2014,Vestergaard2014,Masuda2013a,Rocha2013a} or rigorous theoretical analyses~\cite{Volz2009,Schwarzkopf2010,Perra2012,Taylor2012}. However, there is a lack of methodologies for \emph{containing} epidemic outbreaks on temporal networks (except the work~\cite{Liu2014a} for activity driven networks). This is also the case for adaptive networks. In this latter case, we find in the literature various methods for the analysis of the behavior of spreading processes evolving over adaptively changing temporal networks~\cite{Gross2008,Rogers2012a,Tunc2014,Guo2013,Valdez2012,Wang2011b,Szabo-Solticzky2016} relying on extensive numerical simulations. However, this is a lack of effective control strategies in the context of adaptive networks.

Nevertheless, in recent years, we have witnessed an emerging effort
towards the efficient containment of epidemic processes in temporal and adaptive networks using tools from the field of control theory. The aim of
this chapter is to give an overview of this research thrust by focusing on
optimal resource allocation problems for efficient eradication
of epidemic outbreaks. We specifically focus the scope of this chapter on the following three
classes of temporal and adaptive networks: 1) Markovian temporal
networks~\cite{Ogura2015a}, 2) aggregated-Markovian edge-independent temporal
networks~\cite{Ogura2015c,Nowzari2015b}, and 3) adaptive SIS
models~\cite{Guo2013,Ogura2015i}. We see that the optimal resource allocation
problem in these three cases can be reduced to an efficiently solvable class of
optimization problems called convex programs~\cite{Boyd2004} (more
precisely, geometric programs~\cite{Boyd2007}).

This chapter is organized as follows. After preparing necessary mathematical
notations, in Section~\ref{sec:Markov} we study the optimal resource allocation
problem in Markovian temporal networks. We then focus our exposition on a
specific class of Markovian temporal networks, called aggregated-Markovian
edge-independent temporal networks, in Section~\ref{sec:AMEI}. We finally present
recent results in the context of adaptive SIS models in Section~\ref{sec:ASIS}.

\runinhead{Notation}

We denote the identity matrix by $I$. The maximum real part of the eigenvalues of a square
matrix~$A$ is denoted by $\lambda_{\max}(A)$. For matrices $A_1$, $\dotsc$,
$A_n$, we denote by $\bigoplus_{i=1}^n A_i$ the block-diagonal matrix containing
$A_1$, $\dotsc$, $A_n$ as its diagonal blocks. If the matrices~$A_1$, $\dotsc$,
$A_n$ have the same number of columns, then the matrix obtained by stacking
$A_1$, $\dotsc$, $A_n$ in vertical is denoted by $\col_{i=1}^n A_i$.  An
undirected graph is defined as the pair~$\mathcal{G}=(\mathcal{V},\mathcal{E})$,
where $\mathcal{V}=\{ {1},\dotsc,{n}\} $ is a set of nodes and $\mathcal{E}$ is
a set of edges, defined as unordered pairs of nodes. The adjacency
matrix~$A=[a_{ij}]_{i, j}$ of~$\mathcal{G}$ is defined as the $n\times n$ matrix
such that $a_{ij} = a_{ji} = 1$ if $\{i, j\} \in \mathcal E$, and $a_{ij}=0$
otherwise.

\section{Markovian Temporal Networks} \label{sec:Markov}

Since the dynamics of realistic temporal networks has intrinsic uncertainties
in, for example, the appearance/disappearance of edges, the durations of
temporal interactions, and inter-event times, most mathematical models of
temporal networks in the literature have been written in terms of stochastic
processes. In particular, many stochastic models of temporal networks
(see, e.g., \cite{Perra2012,Volz2009,Clementi2008,Karsai2014}) employ
\emph{Markov processes} due to their simplicity, including time-homogeneity and
memoryless properties. The aim of this section is to present a rigorous and
tractable framework for the analysis and control of epidemic outbreaks taking
place in Markovian temporal networks. We remark that, throughout this chapter,
we shall focus on the specific type of spreading processes described by the
networked SIS model among other networked epidemic models~(see,
e.g.,~\cite{Pastor-Satorras2015a}).

\subsection{Model}\label{sec:model:Markovian}

In this subsection, we present the model of disease spread and temporal networks
studied in this section. We start our exposition from reviewing a model of
spreading processes over static networks called the \emph{Heterogeneous
Networked SIS} (HeNeSIS) model~\cite{Preciado2014}, which is an extension of the
popular $N$-intertwined SIS model~\cite{VanMieghem2009a} to the case of nodes
with heterogeneous spreading rates. Let $\mathcal G$ be an undirected graph having $n$
nodes, where nodes in~$\mathcal G$ represent individuals and edges represent
interactions between them. At a given time~$t \geq 0$, each node can be in one
of two possible states: {\it susceptible} or {\it infected}. In the HeNeSIS
model, when a node~$i$ is infected, it can randomly transition to the
susceptible state with an instantaneous rate~$\delta_i > 0$, called the {\it
recovery rate} of node~$i$. On the other hand, if a neighbor of node~$i$ is in
the infected state, then the neighbor can infect node~$i$ with the instantaneous
rate~$\beta_i$, where $\beta_i > 0$ is called the {\it infection rate} of node
$i$. We define the variable $x_i(t)$ as $x_i(t)=1$ if node~$i$ is infected at
time $t$, and $x_i(t)=0$ if $i$ is susceptible; then, the transition
probabilities of the HeNeSIS model in the time window $[t, t+h]$ can be written
as
\begin{equation}\label{eq:staticSIS}
\begin{aligned}
\Pr(x_i(t+h) = 1 \mid x_i(t) = 0) 
&=
\beta_i \sum_{{j \in \mathcal N_i}} x_j(t) h + o(h),
\\
\Pr(x_i(t+h) = 0 \mid x_i(t) = 1) 
&= 
\delta_i h + o(h), 
\end{aligned}
\end{equation}
where $\mathcal N_i$ is the set of neighbors of node~$i$ and $o(h)/h\to 0$  as
$h \to 0$.

Although the collection of variables $(x_1, \dotsc, x_n)$ is simply a Markov
process, this process presents a total of~$2^n$ possible states (two
states per node). Therefore, its analysis is very hard for arbitrary contact
networks of large size. A popular approach to simplify the analysis of this
type of Markov processes is to consider upper-bounding linear models, as described below. Let $A$
denote the adjacency matrix of~$\mathcal G$. Define the vector $p = (p_1,
\dotsc, p_n)^\top$ and the diagonal matrices $P = \bigoplus(p_1, \dotsc, p_n)$,
$B = \bigoplus(\beta_1, \dotsc, \beta_n)$, and $D = \bigoplus(\delta_1, \dotsc,
\delta_n)$. Then, it is known~\cite{Preciado2014} that the solutions~$p_i(t)$
($i=1$, $\dotsc$, $n$) of the vectorial linear differential equation
\begin{equation}\label{LinearMFA}
dp/dt = (B A-D)p 
\end{equation}
upper-bound the evolution of the infection probabilities $\Pr(\text{$i$ is infected at time $t$})$ from
the exact Markov process with $2^n$ states. Thus, if the solution of
\eqref{LinearMFA} satisfies $p(t) \to 0$ exponentially fast as $t\to \infty$,
then the infection dies out in the exact Markov process exponentially fast as
well. Since the differential equation~\eqref{LinearMFA} is a linear system, the maximum real eigenvalue $\lambda_{\max}(BA-D)$ of the
matrix~$BA-D$ determines the asymptotic behavior of the solution. The above
considerations show that the spreading process dies out exponentially fast if
\begin{equation}\label{eq:threshold:static}
\lambda_{\max}(BA-D) < 0. 
\end{equation}
In the special case of homogeneous infection and recovery rates, i.e., $\beta_i
= \beta$ and $\delta_i = \delta$ for all nodes~$i$,
condition~\eqref{eq:threshold:static} yields the following well-known extinction
condition (see, e.g.,~\cite{Lajmanovich1976,Ahn2013})
\begin{equation}\label{LinearMFA:TI}
\frac{\beta}{\delta} < \frac{1}{\lambda_{\max}(A)}. 
\end{equation}

\begin{figure}[tb]
\centering
\includegraphics[width=.45\linewidth]{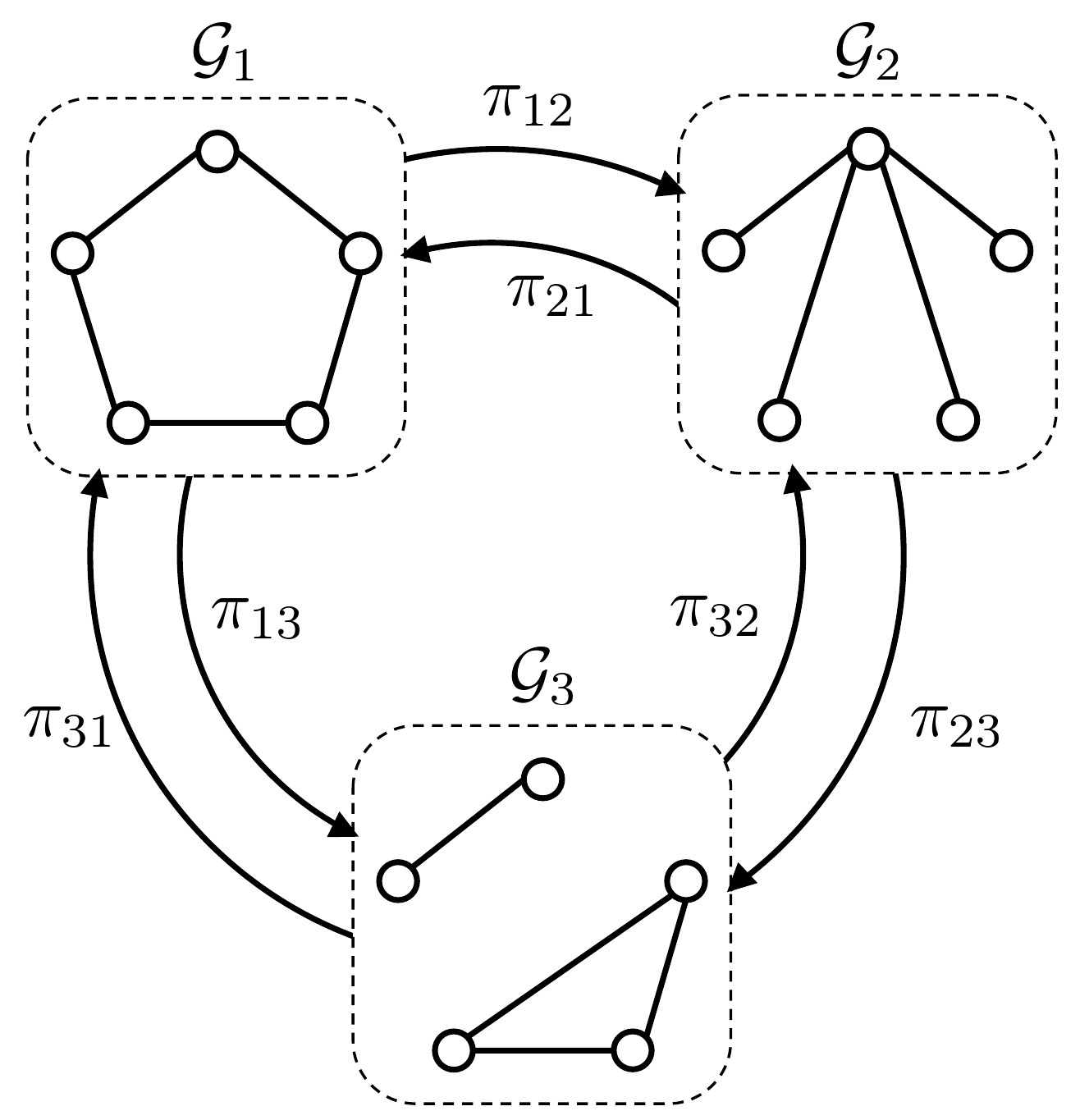}
\caption{Markovian temporal network having three possible configurations ($\mathcal G_1$, $\mathcal G_2$, and $\mathcal G_3$), and the corresponding stochastic transition rates.}
\label{fig:MTN}
\end{figure}

However, conditions \eqref{eq:threshold:static} and
\eqref{LinearMFA:TI} are not applicable to the case of temporal networks having
time-varying adjacency matrices. In this section, we focus on the case where the
dynamics of the temporal network is modeled by a Markov process. In order to
specify a Markovian temporal network, we need the following two ingredients. The
first one is the set of `graph configurations' that can be taken by the temporal
network of our interest. Let those configurations (static and undirected
networks) be $\mathcal G_1$, $\dotsc$, $\mathcal G_L$. This implies that, at
each time $t\geq 0$, the temporal network to be modeled always takes one of the
configurations~$\mathcal G_1$, $\dotsc$, $\mathcal G_L$. The other is the set of
stochastic transition rates between graph configurations. Specifically, we let
$\pi_{k\ell}$ denote the stochastic transition rate from the configuration $\mathcal G_k$
to $\mathcal G_\ell$. This implies that, if the configuration of the temporal
network at time $t$ is $\mathcal G_k$, then the probability of the temporal
network having another configuration $\mathcal G_\ell$ at time $t+h$ equals
$\pi_{k\ell}h + o(h)$. We show a schematic diagram of a Markovian temporal
network in Fig.~\ref{fig:MTN}.

We now describe the model of disease spread considered in this section. Let
$\mathcal G(t)$ be a Markovian temporal network. Let $\mathcal N_i(t)$ be the
set of neighbors of node~$i$ at time $t$ in the graph $\mathcal G(t)$. Then, we can reformulate the
transition probabilities \eqref{eq:staticSIS} of the HeNeSIS model as
\begin{equation}\label{eq:TVSIS}
\begin{aligned}
\Pr(x_i(t+h) = 1 \mid x_i(t) = 0) 
&=
\beta_i \sum_{j \in \mathcal N_i(t)} x_j(t) h + o(h),
\\
\Pr(x_i(t+h) = 0 \mid x_i(t) = 1) 
&= 
\delta_i h + o(h). 
\end{aligned}
\end{equation}
Notice that, in the first equation, the infection probability is dependent not
only on the infection states of the other nodes but also the connectivity of the
network. Then, we can formulate an upper-bounding model for the HeNeSIS model over the
Markovian temporal network $\mathcal G(t)$ as
\begin{equation}\label{eq:upperBoundMarkov}
dp/dt = \left(B A(t) -D\right)p(t), 
\end{equation}
where $A(t)$ denotes the adjacency matrix of~$\mathcal G(t)$.

\subsection{Optimal Resource Distribution} \label{sec:ORD:Markovian}

Let us consider the following epidemiological problem~\cite{Preciado2014}:
Assume that we have access to vaccines that can be used to reduce the infection
rates of individuals in the network, as well as antidotes that can be used to
increase their recovery rates. Assuming that both vaccines and antidotes have an
associated cost and that we are given a fixed budget, how should we distribute vaccines
and antidotes throughout the individuals in the network in order to eradicate an
epidemic outbreak at the maximum decay rate? In what follows, we state this question in rigorous terms and present an optimal solution using an efficient optimization
framework called geometric programming~\cite{Boyd2007}.

Assume that we have to pay $f(\beta_i)$ unit of cost to tune the infection rate
of node~$i$ to $\beta_i$. Likewise, we assume that the cost for tuning the
recovery rate of node~$i$ to $\delta_i$ equals $g(\delta_i)$.
Notice that the total cost of tuning the collection of infection rates
$(\beta_1, \dotsc, \beta_n)$ and recovery rates $(\delta_1, \dotsc, \delta_n)$
in the network  is given by 
\begin{equation*}
R = \sum_{i=1}^n (f(\beta_i) + g(\delta_i)).
\end{equation*}
We further assume that these rates can be tuned within the following feasibility
intervals:
\begin{equation}\label{eq:gammabetabounds}
0< \ubar \beta_i \leq \beta_i \leq \bar \beta_i, \quad 
0< \ubar \delta_i \leq
\delta_i \leq \bar \delta_i. 
\end{equation}
We can now state our optimal resource allocation problem as follows: 

\begin{problem}\label{prb:epidemiology:Markov}
Consider a HeNeSIS spreading process over a Markovian temporal network. Given a
budget $\bar R > 0$, tune the infection and recovery rates~$\beta_i$ and
$\delta_i$ in the network in such a way that the exponential decay rate of the
infection probabilities is maximized while satisfying the budget constraint~$R
\leq \bar R$ and the box constraints~\eqref{eq:gammabetabounds}.
\end{problem}

In order to solve this problem, we first present an analytical framework for
quantifying the decay rate of the infection probabilities, given the parameters
in the HeNeSIS model and the Markovian temporal network. In fact, using tools from control theory, it is possible to prove the following upper-bound on the decay rate of infection probabilities in the HeNeSIS model:

\begin{proposition}\label{prop:analysis:Makorv}
Consider the HeNeSIS spreading process over a Markovian temporal network. Let
$\pi_{\ell\ell} = -\sum_{\ell \neq k}\pi_{\ell k}$. If 
\begin{equation*}
\lambda_{\max}(\mathcal A_1) < 0 
\end{equation*}
for the matrix 
\begin{equation*}
\mathcal A_1 = \begin{bmatrix}
BA_1-D+\pi_{11}I & \ \pi_{21} I\  & \ \cdots\  & \pi_{L1} I
\\
\rule{0pt}{4ex} \pi_{12}I &\ \ddots\  &\ \ddots\  & \vdots
\\
\rule{0pt}{4ex} \vdots &\ \ddots\ &\ \ddots\ & \pi_{L,L-1}I
\\
\rule{0pt}{4ex} \pi_{1L} I &\ \cdots\ &\ \pi_{L-1,L}I\ & BA_L-D + \pi_{LL} I
\end{bmatrix}, 
\end{equation*}
then the infection probabilities of nodes converge to zero exponentially fast
with an exponential decay rate of~$|\lambda_{\max}(\mathcal A_1)|$.
\end{proposition}

Besides providing an analytical method for quantifying the rate of convergence
to the disease-free state, this proposition allows us to sub-optimally minimize
the decay rate of the epidemic outbreak by minimizing the maximum real
eigenvalue~$\lambda_{\max}(\mathcal A_1)$ of the Metzler matrix~$\mathcal A_1$.
In fact, by employing the celebrated Perron-Frobenius theory~\cite{Horn1990} for
nonnegative matrices, we are able to solve Problem~\ref{prb:epidemiology:Markov}
via a class of optimization problems called geometric
programs~\cite[Proposition~10]{Preciado2014}, briefly reviewed
below~\cite{Boyd2007}. Let $x_1$, $\dotsc$, $x_n$ denote positive variables and
define $x = (x_1, \dotsc, x_n)$. In the context of geometric programming, a real function~$g(x)$ is a {\it
monomial} if there exist a $c \geq 0$ and $a_1, \dotsc, a_n \in \mathbb{R}$ such
that $g(x) = c x_{\mathstrut 1}^{a_{1}} \dotsm x_{\mathstrut n}^{a_n}$. Also, we
say that a function~$f(x)$ is a {\it posynomial} if it is a sum of monomials
of~$x$ (we point the readers to~\cite{Boyd2007} for more details). Given a
collection of posynomials $f_0{(x)}$, $\dotsc$, $f_p{( x)}$ and monomials
$g_1{(x)}$, $\dotsc$, $g_q{(x)}$, the optimization problem
\begin{equation*} 
\begin{aligned}
\minimize_{x>0}\ \ \ 
&
f_0(x)
\\
\subjectto\ \ \ 
&
f_i(x)\leq 1,\quad i=1, \dotsc, p, 
\\
&
g_j(x) = 1,\quad j=1, \dotsc, q, 
\end{aligned}
\end{equation*}
is called a {\it geometric program}. A constraint of the form $f(x)\leq 1$ with
$f(x)$ being a posynomial is called a posynomial constraint. It is
known~\cite{Boyd2007} that a geometric program can be efficiently converted into
an equivalent convex optimization problem, which can be solved in polynomial time~\cite{Boyd2004}.

We can now state the first main result of this chapter: 

\begin{theorem}[{\cite[Section~VI]{Ogura2015a}}]\label{thm:control:Markovian}
Assume that the cost function $f$ is a posynomial and, also, there exists
$\hat\delta > \bar\delta$ such that the function $\tilde g(\tilde \delta) =
g(\hat \delta - \tilde \delta)$ is a posynomial in~$\tilde \delta$. Then, the
infection and recovery rates that solve Problem~\ref{prb:epidemiology:Markov}
are given by $\{\beta_i^\star\}_{i=1}^n$ and $\{\hat \delta - \tilde
\delta_i^\star\}_{i=1}^n$, where the starred variables solve the optimization
problem
\begin{equation}\label{eq:optimization:Markovian}
\begin{aligned}
\minimize_{\beta_i,\,\tilde \delta_i,\,v>0,\,\lambda>0}\ \ \ 
& 
1/\lambda
\\
\text{\upshape subject to}\ \ \ \ \ 
&\mathcal A_1 v \leq -\lambda v, 
\\
& {\sum_{i=1}^n (f(\beta_i) + \tilde g(\tilde \delta_i))}\leq \bar R, 
\\
&  {\ubar{\beta}} \leq \beta_i \leq \bar{\beta}, 
\\
&
\hat{\delta} - \bar{\delta} \leq \tilde{\delta}_i \leq \hat{\delta} - \ubar{\delta}. 
\end{aligned}
\end{equation}
Moreover, this optimization problem can be equivalently converted to a geometric
program.
\end{theorem}

It is rather straightforward to verify that the optimization problem
\eqref{eq:optimization:Markovian} can be converted to a geometric program. For
example, one can easily confirm that the vectorial constraint~$\mathcal A_1 v
\leq -\lambda v$ is equivalent to the following set of posynomial constraints
\begin{equation*}
\dfrac{
(\sum_{k\neq \ell} \pi_{k\ell}v_{ki})
+
\beta_i \sum_{j=1}^n [A_\ell]_{ij} v_{\ell j}+
\tilde \delta_i v_{\ell i}+\lambda v_{\ell i}
}{
(\hat \delta_i -\pi_{\ell})v_{\ell i}
} \leq 1,
\end{equation*}
for all $i=1$, $\dotsc$, $n$ and $\ell=1$, $\dotsc$, $L$. We refer the
interested readers to the references~\cite{Preciado2014,Ogura2015a} for details.

\subsection{Numerical Simulations}\label{sec:simulation:Markov}

\begin{figure}[tb]
\centering
\includegraphics[width=1\linewidth]{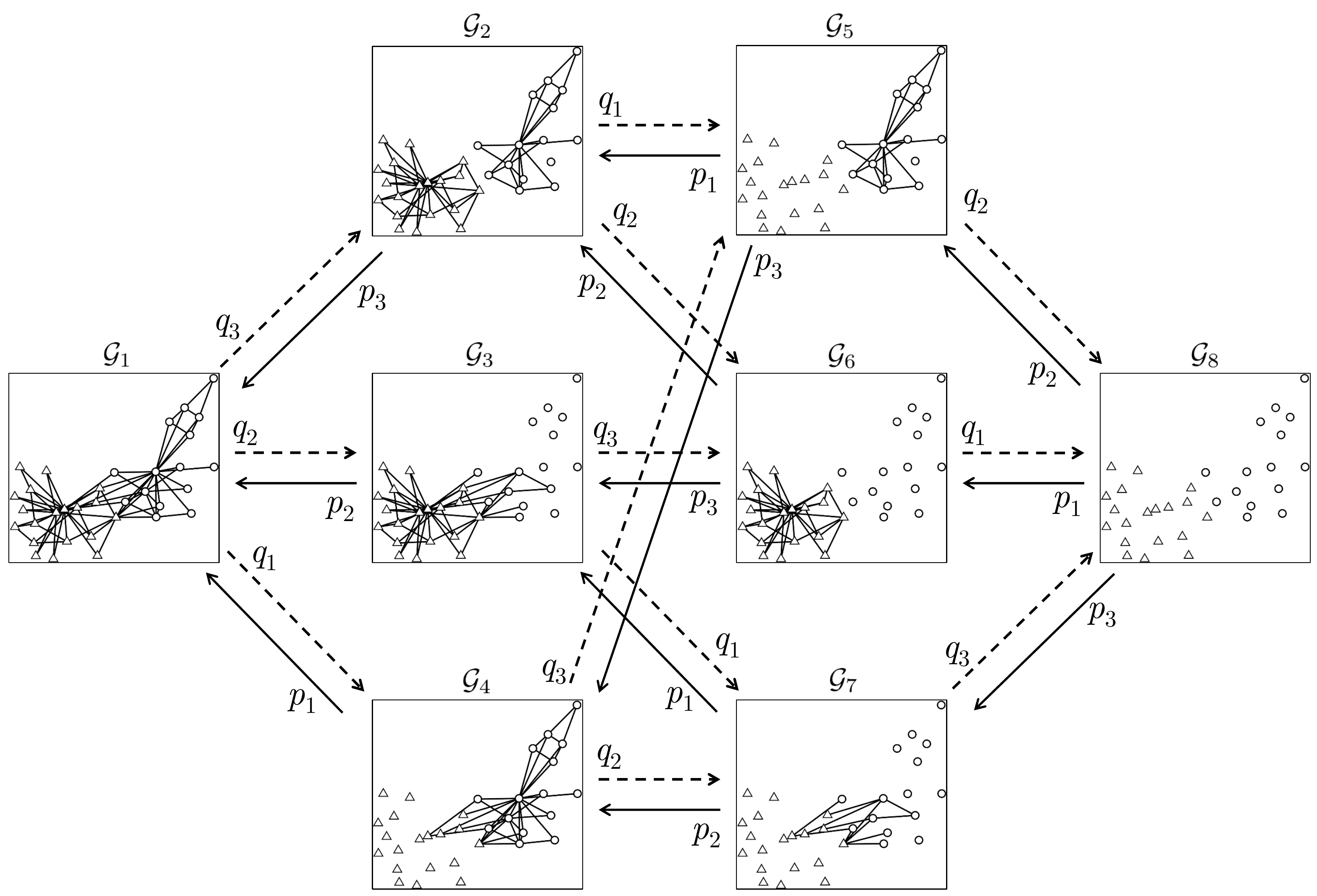}
\caption{Transition diagram of the Markovian Karate Club Network}
\label{fig:KarateMarkovianNet}
\end{figure}

To illustrate the results presented in this section, we consider the HeNeSIS
model over the following Markovian temporal network based on the well-known Zachary Karate Club network~\cite{Zachary1977}. In order to construct
a Markovian temporal network from this static network, we first identify two
clusters (i.e., a division of the set of nodes into two disjoint subsets) in the
network using the spectral clustering technique~(see, e.g.,
\cite{VonLuxburg2007}). We then classify the edges in the static network into
the following three classes: edges within the first cluster ($\mathcal
E^{(1)}$), within the second cluster ($\mathcal E^{(2)}$), and between distinct
clusters ($\mathcal E^{(3)}$). We then consider the following stochastic temporal behavior for the network structure: edges in each class $\mathcal
E^{(i)}$ ($i=1, 2, 3$) appear or disappear simultaneously, with an activation
rate~$p_i$ and a deactivation rate~$q_i$, respectively. Notice that, in this
setting, the temporal network has a total of~$2^3 = 8$ configurations $\mathcal
G_\ell = (\mathcal V, \mathcal E_\ell)$ ($\ell=1$, $\dotsc$, $8$) having the
sets of edges listed below:
\begin{equation*}
\begin{aligned}
&&\mathcal E_1 &= \mathcal E^{(1)} \cup \mathcal E^{(2)} \cup \mathcal E^{(3)}, &&
\\
\mathcal E_2 &= \mathcal E^{(1)} \cup \mathcal E^{(2)}, 
&\mathcal E_3 &= \mathcal E^{(2)} \cup \mathcal E^{(3)}, 
&\mathcal E_4 &= \mathcal E^{(1)}  \cup \mathcal E^{(3)}, 
\\
\mathcal E_5 &= \mathcal E^{(1)}, 
&\mathcal E_6 &= \mathcal E^{(2)}, 
&\mathcal E_7 &= \mathcal E^{(3)}, 
\\
&&\mathcal E_8 &= \emptyset. &&
\end{aligned}
\end{equation*}
We show the transition diagram of the resulting Markovian temporal network
(called the Markovian Karate Network) in Fig.~\ref{fig:KarateMarkovianNet},
where solid (dashed) arrows indicate transitions involving the activation
(deactivation, respectively) of edges.

\begin{figure}[tb]
\centering
\includegraphics[width=.5\linewidth]{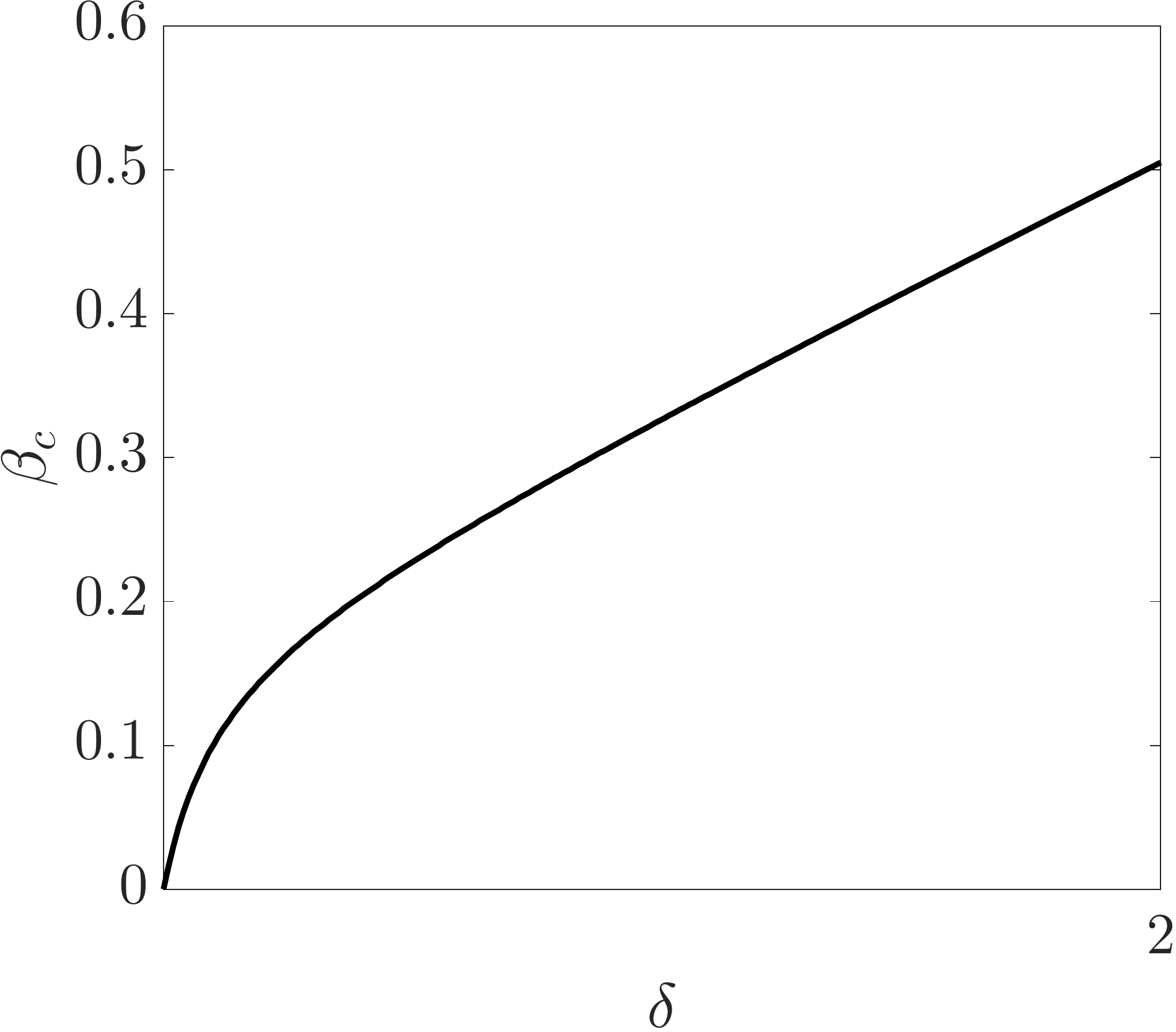}
\caption{The epidemic threshold $\beta_c$ of the Markovian Karate Club Network versus the recover rate~$\delta$.}
\label{fig:threshold:Markovian}
\end{figure}

Using Proposition~\ref{prop:analysis:Makorv}, we first illustrate how the
time-variability of the Markovian Karate Network affects the behavior of
epidemic threshold. We let the activation and deactivation rates of the edges
be $p_1 = p_2 = 0.1$, $q_1 = q_2 = 1$, $p_3 = 0.02$, and $q_3 = 5$. As for the
HeNeSIS model, we vary the value of the recovery rate~$\delta$, uniformly at random among
nodes, from $0$ to $2$. For each value of~$\delta$, we use a bisection search to
find the supremum $\beta_c$ of the transmission rate~$\beta$ that guarantees the
exponentially fast extinction of the disease spread (i.e.,
$\lambda_{\max}(\mathcal A_1) < 0$). We show the obtained values
of~$\beta_c$ versus $\delta$ in Fig.~\ref{fig:threshold:Markovian}. We can
observe that, unlike in the case of static network, the threshold value
$\beta_c$ in our case exhibits a nonlinear dependence on~$\delta$.

\begin{figure}[tb]
\centering
\includegraphics[width=.3\linewidth]{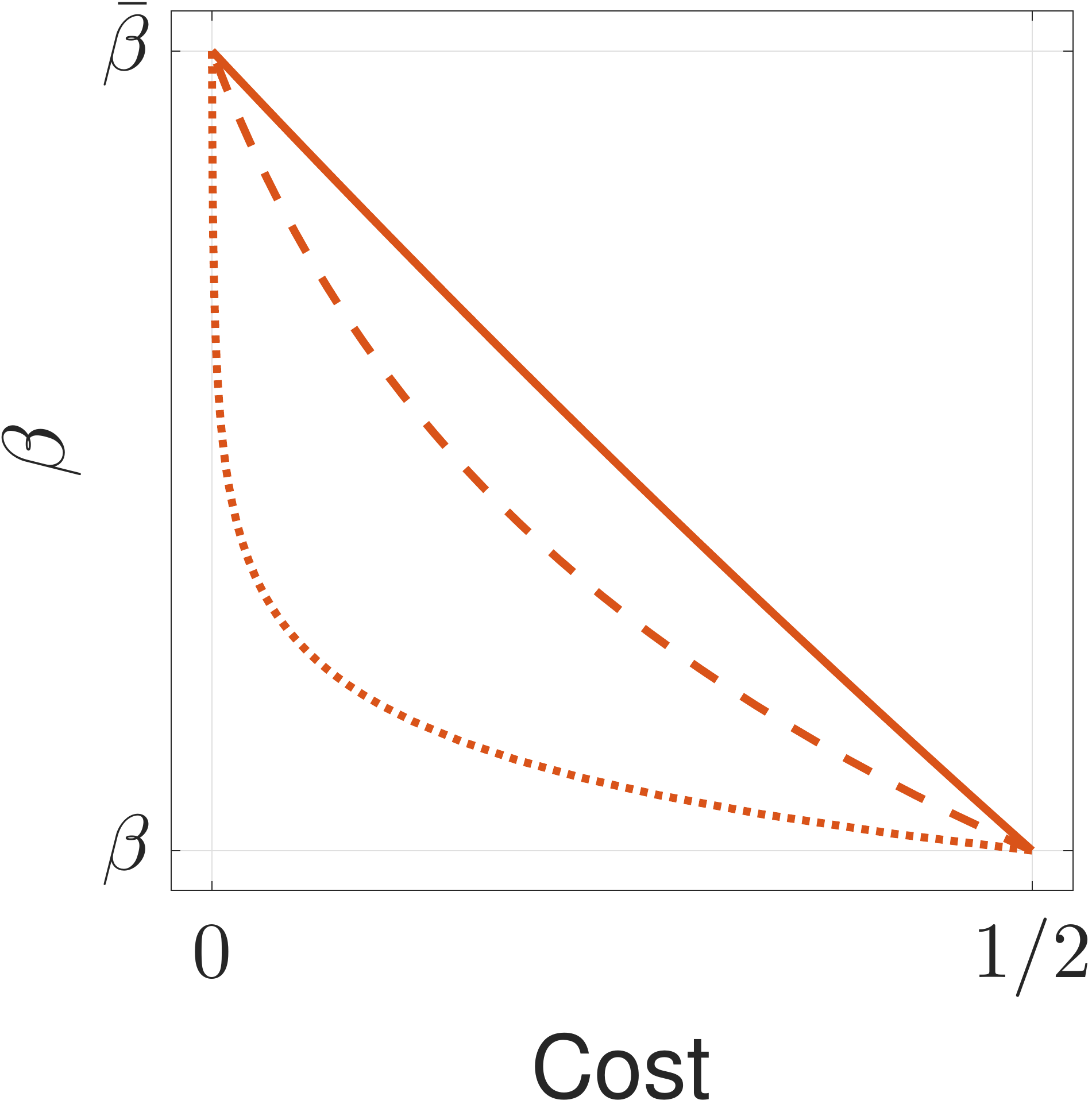}
\hspace{.05\linewidth}
\includegraphics[width=.3\linewidth]{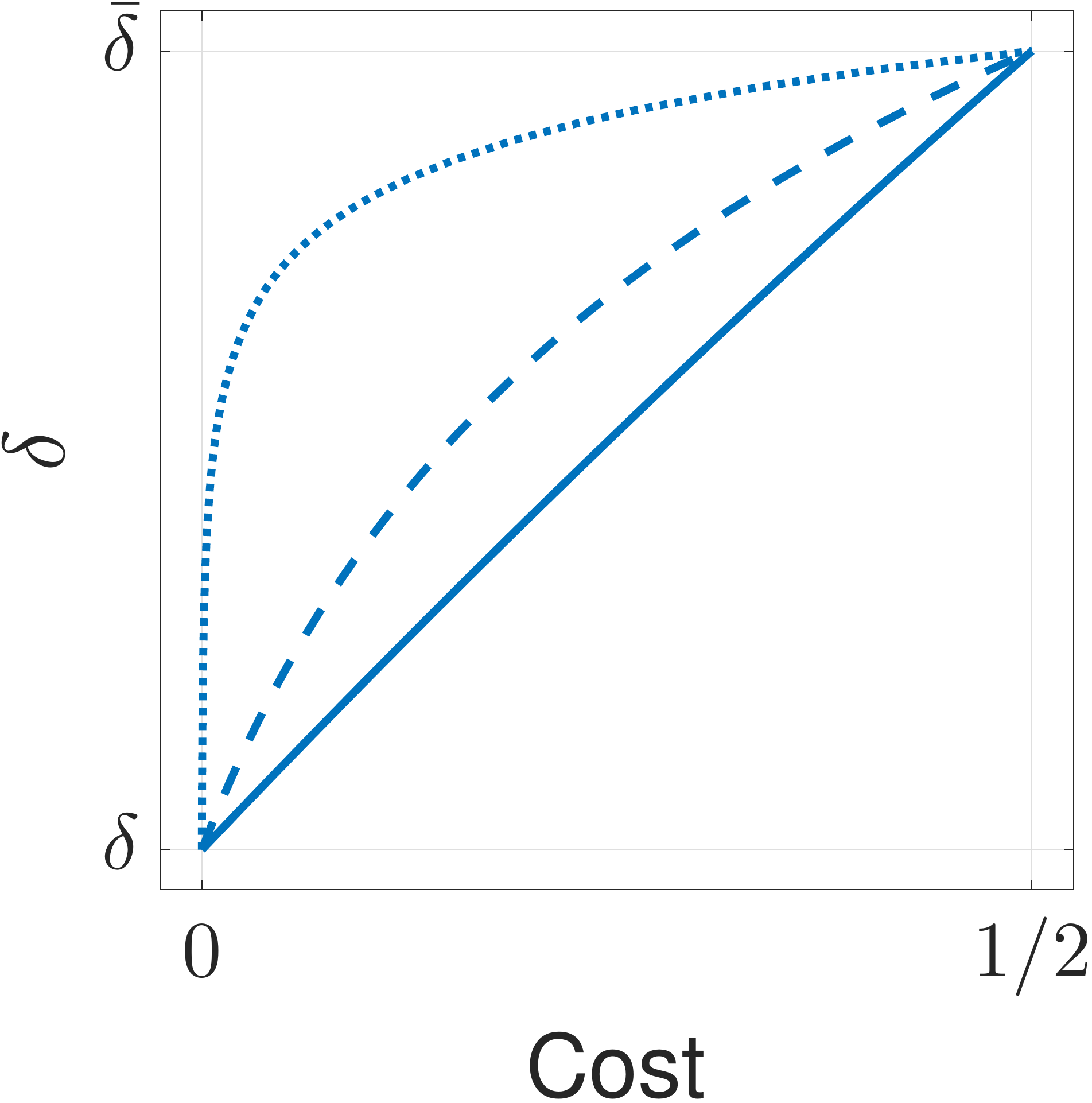}
\caption{Left: Cost functions for transmission rates. Solid: $q = 0.1$, dashed: $q = 10$, dotted: $q = 50$. Right: Cost function for recovery rates. $g$. Solid: $r = 0.1$, dashed: $r = 10$, dotted: $r = 50$.}
\label{fig:costFunctions}
\end{figure} 

\begin{table*}[tb]
\centering
\newcommand{\gwidth}{.15\textwidth}
\newcommand{\gheight}{4.3cm}
\caption{Optimal investments for a) infection rates and b) recovery rates in the case of 1) the Markovian network and 2) time-aggregated network, respectively. Darker node colors represent heavier investments to vaccinate/antidote the node, while white nodes do not receive any investment.}
\label{table:optimalInvestments}
\begin{tabular}{ccc}
\hline	
& a) Infection rates & b) Recovery rates 
\\
\hhline{===}
\parbox[c][\gheight]{\gwidth}{1) Markovian\\\phantom{1) }network} &
\parbox[c]{\resourceGraphWidth}{\includegraphics[width=\resourceGraphWidth]{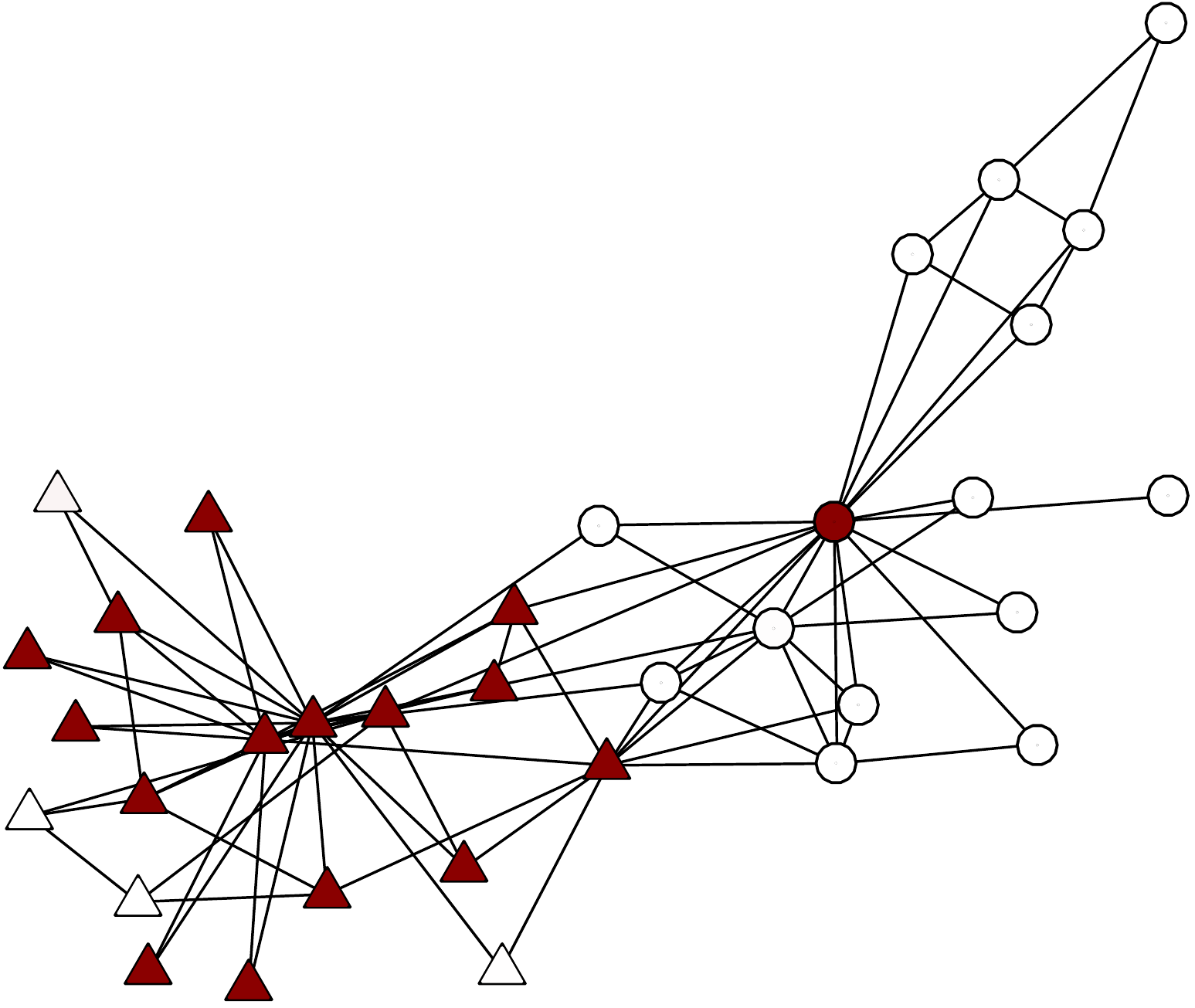}}&
\parbox[c]{\resourceGraphWidth}{\includegraphics[width=\resourceGraphWidth]{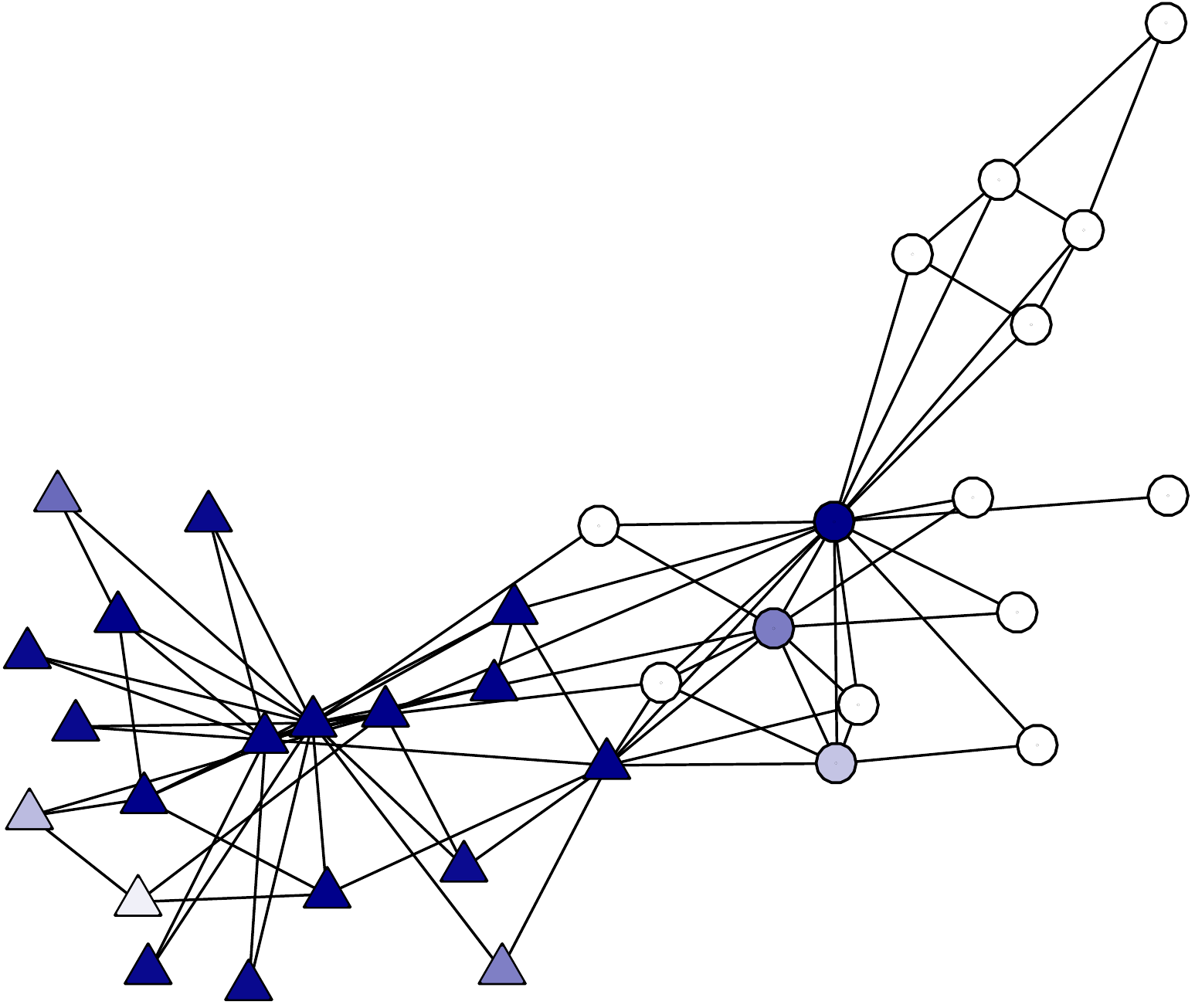}}\\
\hline
\parbox[c][\gheight]{\gwidth}{2) Time-\\\phantom{2) }aggregated\\\phantom{2) }network} &
\parbox[c]{\resourceGraphWidth}{\includegraphics[width=\resourceGraphWidth]{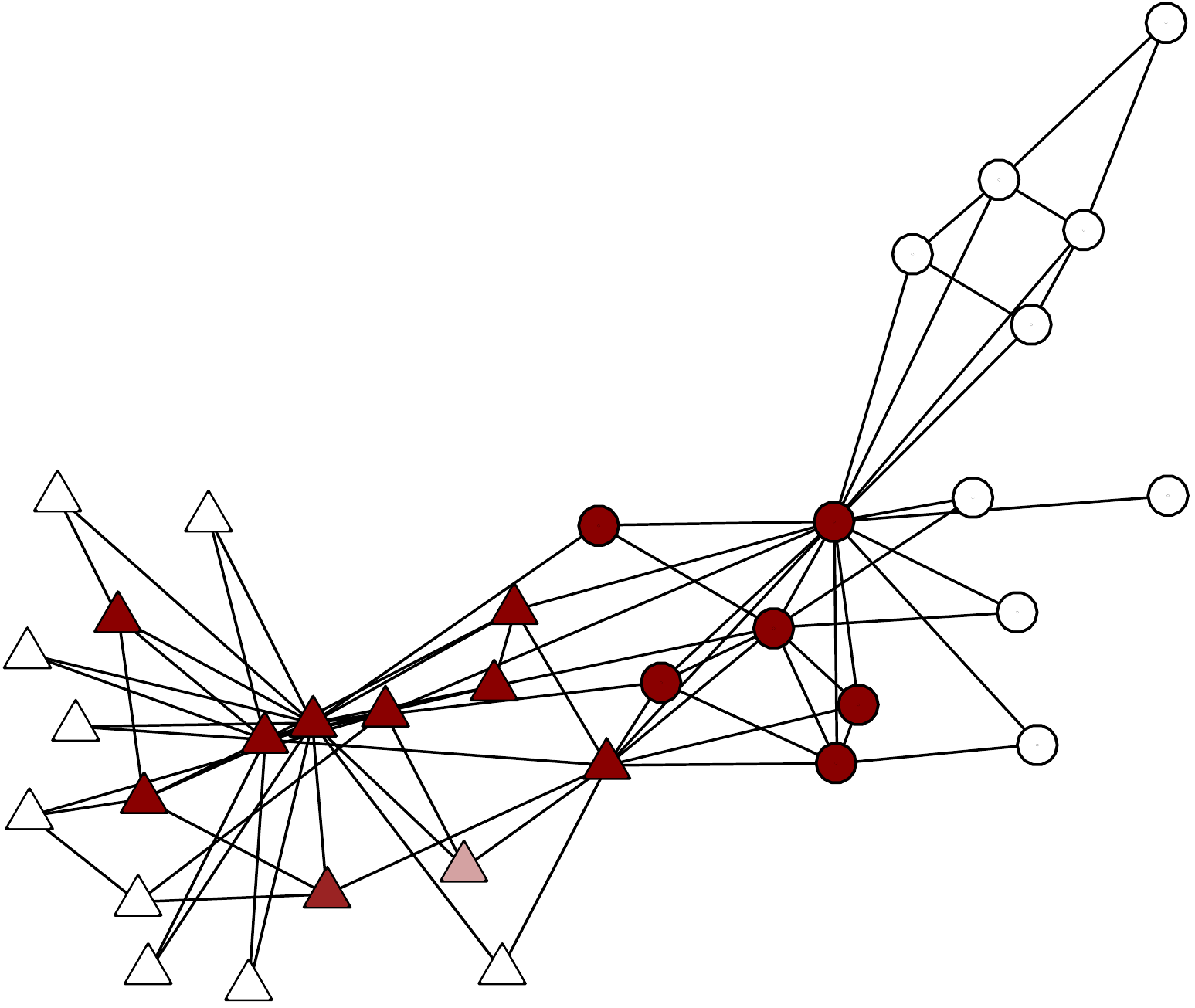}}&
\parbox[c]{\resourceGraphWidth}{\includegraphics[width=\resourceGraphWidth]{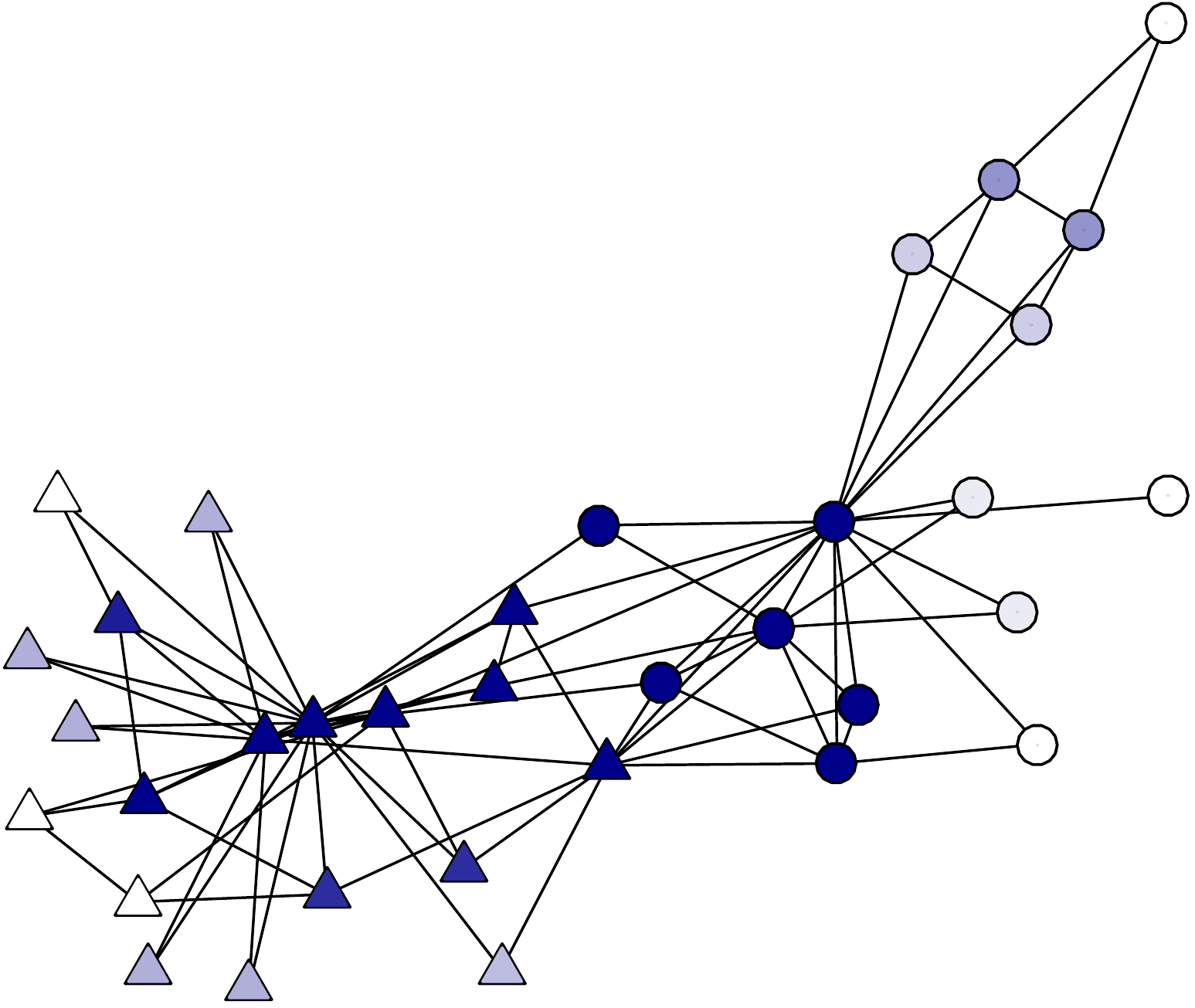}}
\\
\hline
\end{tabular}
\end{table*}

We then move to the cost-optimal eradication of epidemic outbreaks over the
Markovian Karate Network. Let us fix $\ubar \delta = p_1/2 = 0.05$ and $\bar
\beta = \beta_c$, which are considered to be the `natural' recovery and
infection rates of the nodes. We then assume that a full dose of vaccinations
and antidotes can improve these rates at most 20\%, i.e., we let
\begin{equation*}
\ubar \beta = ({0.8})
\bar\beta,\quad \bar \delta = (1.2)\ubar \delta.
\end{equation*} 
The cost functions for tuning the rates are set to be
\begin{equation}\label{eq:costFunctions}
f(\beta) = c_1+c_2/{\beta^q},
\quad 
g(\delta) = c_3+c_4/{(\hat \delta - \delta)^r}, 
\end{equation}
where $q$ and $r$ are positive parameters that allow us to tune the shape of the cost
functions; $c_1$, $\dotsc$, $c_4$ are constants to normalize the cost functions
in such a way that $f(\ubar \beta) = 1/2$, $f(\bar \beta) = 0$, $g(\ubar \delta)
= 0$, and $g(\ubar \delta) = 1/2$. Notice that, with this choice of the
normalization constants, we have $R = 0$ if $(\beta_i, \delta_i) = (\bar \beta,
\ubar \delta)$ for every node~$i$ (i.e., all nodes keep their natural infection
and transmission rates), while $R=n$ (full protection) if $(\beta_i, \delta_i) =
(\ubar \beta, \bar \delta)$ for every $i$ (i.e., all nodes receive the full
amount of vaccinations and antidotes). We show plots of the above cost
functions for various values of~$q$ and $r$ in Fig.~\ref{fig:costFunctions},
when $\hat \delta =2\bar \delta$. In our numerical simulation, we use the
values~$q = r = 0.1$, in which case the cost functions becomes almost linear
(solid lines in Fig.~\ref{fig:costFunctions}). Setting the available budget as
$\bar R = n/2$, we solve the optimization problem in
Theorem~\ref{thm:control:Markovian} and obtain the sub-optimal resource
allocation over the Markovian Karate Network (illustrated in the first row of
Table~\ref{table:optimalInvestments}). We can observe that the nodes at the
`boundaries' of clusters do not receive much investments. This is reasonable
because the first and second clusters are effectively disconnected (due to the
low activation rate~$p_3$ and the high deactivation rate~$q_3$ of edges between
clusters) and, therefore, we do not need to worry too much about the infections
occurring across different clusters.

For the sake of comparison, we solve the same resource allocation problem
for the original (static) Karate Club Network using the framework presented
in~\cite{Preciado2014} and obtain another allocation of vaccines and antidotes over the network
(shown in the second row of Table~\ref{table:optimalInvestments}). We can see
that, in the latter allocation of resources, some of the nodes at the boundaries of the
clusters receive a full investment, unlike in the Markovian case. This
observation shows that, by taking into account the time-variability of temporal
networks, we are able to distribute resources in a more efficient manner.

\section{Edge-Independent Networks} \label{sec:AMEI}

Although the framework presented in the previous section can theoretically deal
with epidemic outbreaks on temporal networks presenting the Markovian property, the framework is not
necessarily applicable to some realistic temporal networks having a large number
of graph configurations (i.e., when the number $L$ is large under the notation
in Section~\ref{sec:model:Markovian}). For example, in the example studied in
Section~\ref{sec:simulation:Markov}, it would be more realistic to assume that
the activations and deactivations of edges within a cluster or between clusters
occur not simultaneously (as assumed in the example) but rather
\emph{respectively} (or, \emph{independently} of each other). However, if we
allow independent edge activations and deactivations for all the 78 edges in the
network, we would end up obtaining a Markovian temporal network having $L =
2^{78} > 10^{23}$ possible graph configurations, which makes  the optimization problem~\eqref{eq:optimization:Markovian} computationally
hard to solve.

The aim of this section is to present an optimization framework to contain
epidemic outbreaks over temporal networks where edges are allowed to activate
and deactivate independently of each other. We specifically focus on the HeNeSIS
model evolving over \emph{aggregated-Markovian edge-independent} (\emph{AMEI})
temporal networks introduced in~\cite{Ogura2015c}. We present an efficient
method for sub-optimally tuning the infection and recovery rates of the nodes in
the network for containing epidemic outbreak in AMEI temporal networks. Unlike
the optimization problem~\eqref{eq:optimization:Markovian}, the computational
complexity for solving the optimization problem presented in this section
\emph{does not} grow with respect to the number $L$ of graph configurations. We
also remark that another advantage of the AMEI temporal networks is its ability
of modeling non-exponential, heavy-tail distributions of inter-event times found
in several experimental studies~\cite{Cattuto2010,Stehle2011}.

\subsection{Model}

We start by presenting the definition of aggregated-Markovian
edge-independent (AMEI) temporal network model~\cite{Ogura2015c}. For simplicity
in our exposition, we shall adopt a formulation slightly simpler than the original
one in~\cite{Ogura2015c}.

\begin{figure}[tb]
\centering
\includegraphics[width=.925\linewidth]{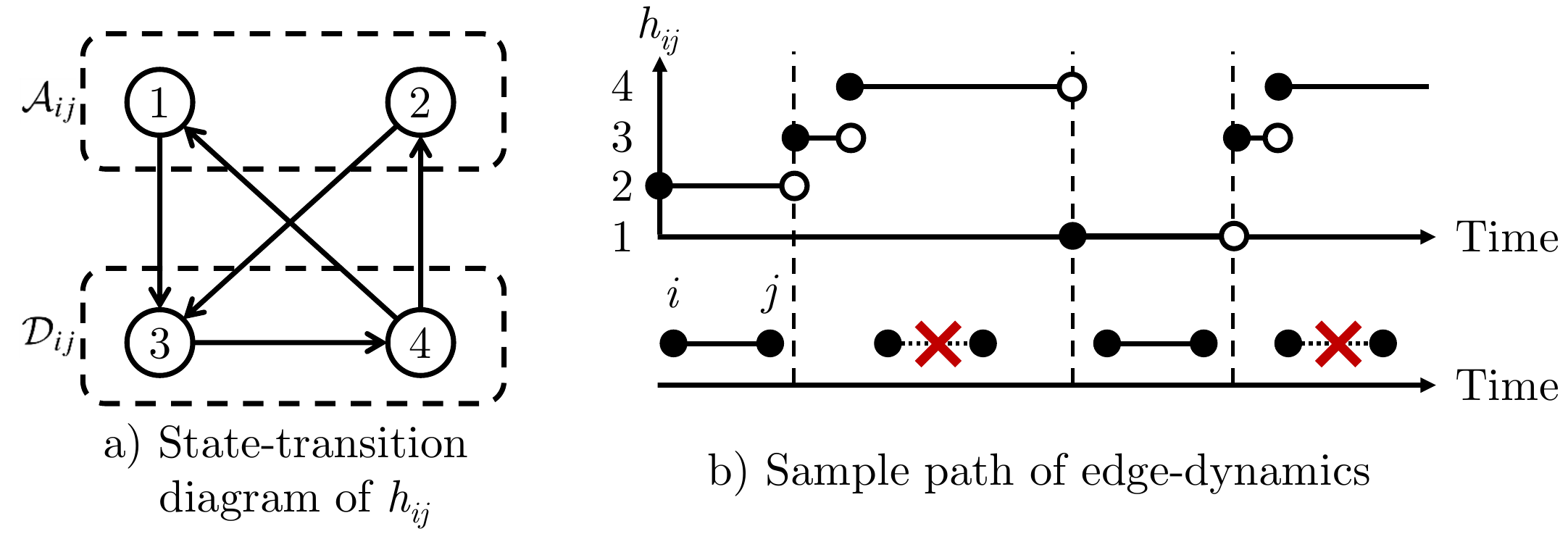}
\caption{Example of the stochastic transitions of a particular edge in an AMEI model. Left: State-transition diagram of the Markov
process~$h_{ij}$ having the state space $\{1, 2 , 3, 4\}$ partitioned into the active set~$\mathcal A_{ij} = \{1, 2\}$ and the inactive set~$\mathcal D_{ij} = \{3, 4\}$. Right: a sample path of
the time-evolution of the edge $\{i, j\}$. } \label{fig:Markov vs AMEI}
\end{figure}

\begin{definition}[{\cite{Ogura2015c}}]
Consider a collection of stochastically independent Markov processes~$h_{ij}$
($1\leq i<j\leq n$) taking values in the integer set~$\{1, \dotsc, M_{ij}\}$. We partition the integer set $M_{ij}$ into $\{1, \dotsc, M_{ij} \}= \mathcal A_{ij} \cup \mathcal
D_{ij}$, where $\mathcal A_{ij}$ is called the active set and $\mathcal D_{ij}$ the inactive set. An \emph{aggregated-Markovian edge-independent} (\emph{AMEI} for short)
temporal network is a random and undirected temporal network~$\mathcal G(t)$ in
which the edge $\{i, j\}$ is present at time $t$ if $h_{ij}(t)$ is in the
\emph{active set}~$\mathcal A_{ij}$ and not present if $h_{ij}(t)$ is in the
\emph{inactive set}~$\mathcal D_{ij}$ (see Fig.~\ref{fig:Markov vs AMEI} for an
illustration).
\end{definition}

A few remarks on the AMEI temporal networks are in order. First, the independence of
the Markov processes $h_{ij}$ for all pairs of nodes ensures the independent
dynamics of the connectivity of any node-pairs, unlike in the example presented
in Section~\ref{sec:simulation:Markov}. Secondly, in the special case where
$M_{ij} = 2$, $\mathcal A_{ij} = \{1\}$, and $\mathcal D_{ij} = \{2\}$ for all
$i$ and $j$, AMEI temporal networks reduce to the well-known model of temporal
networks called the edge-Markovian model~\cite{Clementi2008}. Thirdly, AMEI
temporal networks in fact allow us to model a wider class of temporal networks. For
example, in an edge-Markovian graph, the time it takes for an edge to switch
from connected to disconnected (or vice versa) must follow an exponential
distribution. In contrast, in an AMEI temporal network, we can \emph{design} the
active and inactive sets $\mathcal A_{ij}, \mathcal D_{ij}$ as well as the
Markov process~$h_{ij}$ to fit any desired distribution for the contact
durations with an arbitrary precision~\cite[Example~1]{Ogura2015c}. Finally,
since all the processes~$h_{ij}$ are Markovian, the dynamics of an AMEI temporal
network can be described by the collection~$h = (h_{ij})_{i, j}$, which is again
a Markov process.

\subsection{Optimal Resource Allocation}

In this section, we consider the same epidemiological problem as
Problem~\ref{prb:epidemiology:Markov}:

\begin{problem}[\cite{Nowzari2015b}]\label{prb:epidemiology:AMEI}
Consider a HeNeSIS model over an AMEI temporal network. Given a budget $\bar R >
0$, tune the infection and recovery rates~$\beta_i$ and $\delta_i$ in the
network in such a way that the exponential decay rate of the infection
probabilities is minimized while satisfying the budget constraint~$R \leq \bar
R$ and the box constraints in~\eqref{eq:gammabetabounds}.
\end{problem}

We notice that, although Problem~\ref{prb:epidemiology:AMEI} is a particular
case of Problem~\ref{prb:epidemiology:Markov} for general Markovian temporal
networks since an AMEI temporal network is Markovian, we cannot necessarily
apply the optimization framework presented in
Theorem~\ref{thm:control:Markovian} to the current case. Notice that an AMEI
temporal network allows a total of~$2^m$ graph configurations, where $m =
2^{n(n-1)/2}$ is the number of the undirected edges that can exist in the
network. This implies that the dimension of the vector-valued decision
variable~$v$ in the optimization problem~\eqref{eq:optimization:Markovian}, $nL
= n2^m$, grows exponentially fast with respect to~$n$, making it very hard to
efficiently solve the optimization problem~\eqref{eq:optimization:Markovian}
even for small-scale networks.
We further emphasize that this difficulty cannot be relaxed as long as we rely
on the estimate on the decay rate of infection probabilities presented in
Proposition~\ref{prop:analysis:Makorv}, because the estimate already relies on a
matrix of dimensions $(nL) \times (nL)$. This observation motivates us to derive an
alternative, computationally efficient method for estimating the decay rate of
infection probabilities. In this direction, using tools from random matrix theory, we are able to derive an alternative, tractable
extinction condition for spreading processes over AMEI temporal
networks~\cite{Ogura2015c}:

\begin{proposition}[{\cite[Theorem~3.4]{Ogura2015c}}]\label{prop:analysis:AMEI}
For positive constants~$b$ and $d$, define the decreasing function $\kappa(s) =
n\exp(s/b)[(bs+d)/d]^{-(bs+d)/b^2}$ for $s\geq 0$. Let us consider the HeNeSIS
spreading process over an AMEI temporal network. Define the $n\times n$ matrix~$\bar A =
[\bar a_{ij}]_{i,j}$ by
\begin{equation}\label{eq:def:barA}
\bar a_{ij} = 
\lim_{t\to\infty} \Pr(a_{ij}(t) = 1).
\end{equation}
Let $\Delta = \max_{1\leq i\leq n} \sum_{j=1}^n \left(\beta_i\beta_j \bar
A_{ij}(1-\bar A_{ij}) \right)$ and $c = \eta( B (\sgn \bar A) - D) -
\kappa_{\beta_{\max}, \Delta}^{-1}\!(1)$, where $\sgn(\cdot)$ denotes the
entry-wise application of the sign function and $\beta_{\max} = \max_{1\leq
i\leq n}\beta_i$. 
If 
\begin{equation}\label{eq:threshold:edges}
\lambda_{\max}(\mathcal A_2) < \tau, 
\end{equation}
where 
\begin{equation*}
\mathcal A_2 = B\bar A- D
\end{equation*}
and 
\begin{equation*}
\tau =  \max_{ s \in (\kappa_{\beta_{\max}, \Delta}^{-1}\!(1), \ubar{\delta} + (\abs{c} - c)/2] 
} \left(-\frac{s+c\kappa_{\beta_{\max}, 
\Delta}(s)}{1-\kappa_{\beta_{\max}, \Delta}(s)}\right),
\end{equation*}
then the infection probabilities converge to zero exponentially fast, almost
surely.
\end{proposition}

The extinction condition~\eqref{eq:threshold:edges} is comparable with the
condition in~\eqref{eq:threshold:static} for static networks. Roughly speaking, we can
understand $\bar A$ defined in~\eqref{eq:def:barA} as the adjacency matrix of a
weighted \emph{static} network $\bar{\mathcal G}$ `representing' the original
AMEI temporal network, while $\tau$ can be regarded as a safety margin we have
to take at the cost of utilizing this simplification. We further notice that the
static network~$\bar{\mathcal G}$ arises by taking a long-time limit of the
original AMEI temporal network~\cite{Ogura2015c}. We finally remark that it is
possible to upper-bound the decay rate of the convergence of infection
probabilities using the maximum real eigenvalue of $\mathcal A_2$ (for details,
see \cite{Ogura2015c}).

Proposition~\ref{prop:analysis:AMEI} gives us the following two alternative
options to solve Problem~\ref{prb:epidemiology:AMEI} for a given HeNeSIS
spreading process over an AMEI temporal network: 1) to increase $\tau$ or 2) to
decrease $\lambda_{\max}(\mathcal A_2)$. Among these two options, the former is
not realistic because $\tau$ has a complicated expression and depends on
relevant parameters in a highly complex manner. On the other hand, the maximum
real eigenvalue~$\lambda_{\max}(\mathcal A_2)$ is easily tractable by the
framework used in Section~\ref{sec:ORD:Markovian}. This consideration leads us
to the following sub-optimal solution to Problem~\ref{prb:epidemiology:AMEI}:

\begin{theorem}[{\cite[Section~VI]{Ogura2015a}}]\label{thm:control:AMEI}
Assume that the cost function $f$ is a posynomial and, also, there exists $\hat
\delta > \bar \delta$ such that the function $\tilde g(\tilde \delta) = g(\hat
\delta - \tilde \delta)$ is a posynomial in $\tilde \delta$. Then, the infection
and recovery rates that sub-optimally solve Problem~\ref{prb:epidemiology:AMEI}
are given by $\{\beta_i^\star\}_{i=1}^n$ and $\{\hat \delta - \tilde
\delta_i^\star\}_{i=1}^n$, where the starred variables solve the optimization
problem
\begin{equation*}
\begin{aligned}
\minimize_{\beta_i,\,\tilde \delta_i,\,v>0,\,\lambda>0}\ \ \ 
& 
1/\lambda
\\
\text{\upshape subject to}\ \ \ \ \ 
&\mathcal A_2v \leq -\lambda v, 
\\
& {\sum_{i=1}^n (f(\beta_i) + \tilde g(\tilde \delta_i))}\leq \bar R, 
\\
&  {\ubar{\beta}} \leq \beta_i \leq \bar{\beta}, 
\\
&
\hat{\delta} - \bar{\delta} \leq \tilde{\delta}_i \leq \hat{\delta} - \ubar{\delta}.
\end{aligned}
\end{equation*}
Moreover, this optimization problem can be equivalently converted to a geometric program. 
\end{theorem}

\subsection{Numerical simulation}

\begin{figure}[tb]
\centering
\includegraphics[width=\resourceGraphWidth]{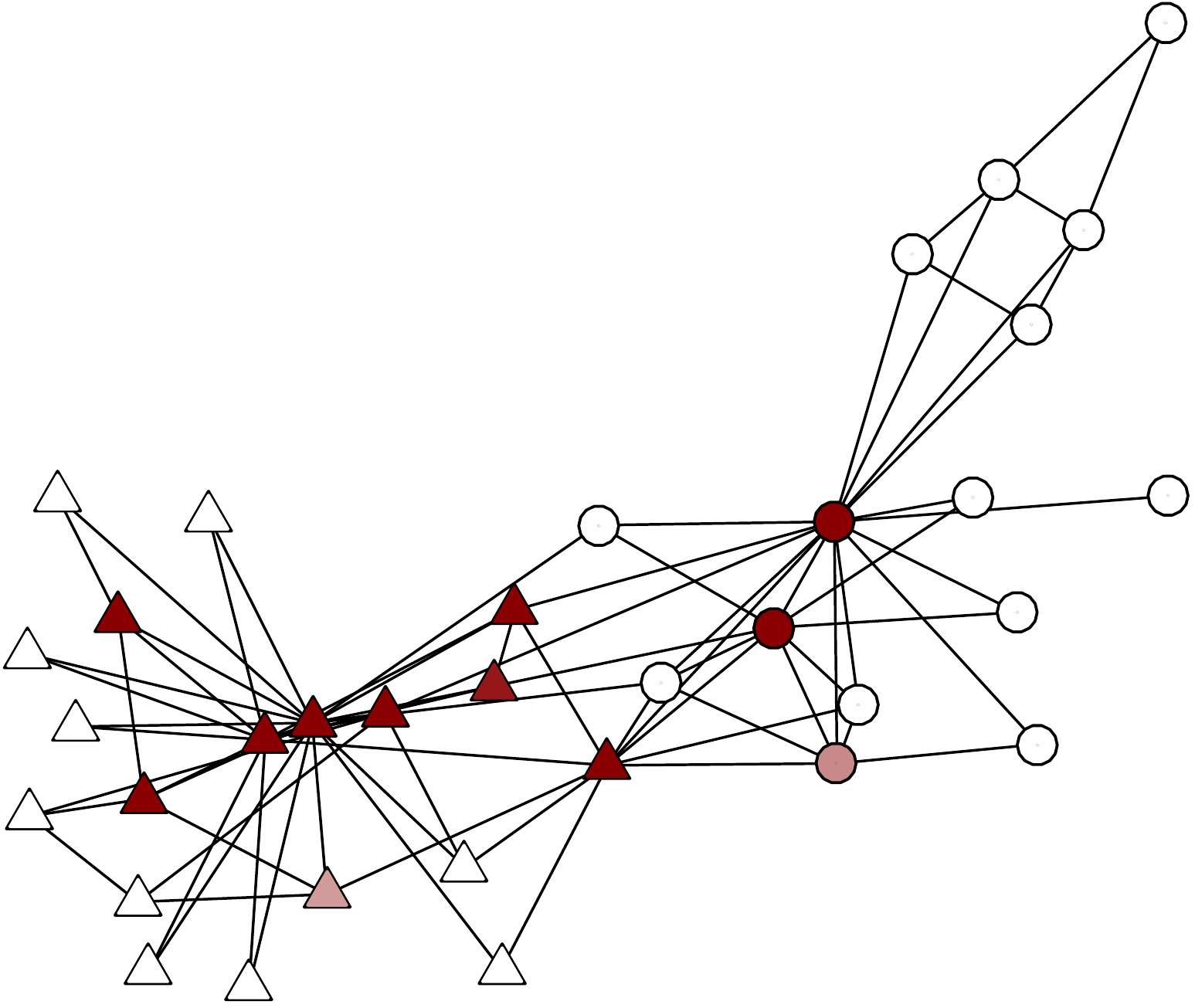}
\hspace{\betweenResourceGraphWidth}
\includegraphics[width=\resourceGraphWidth]{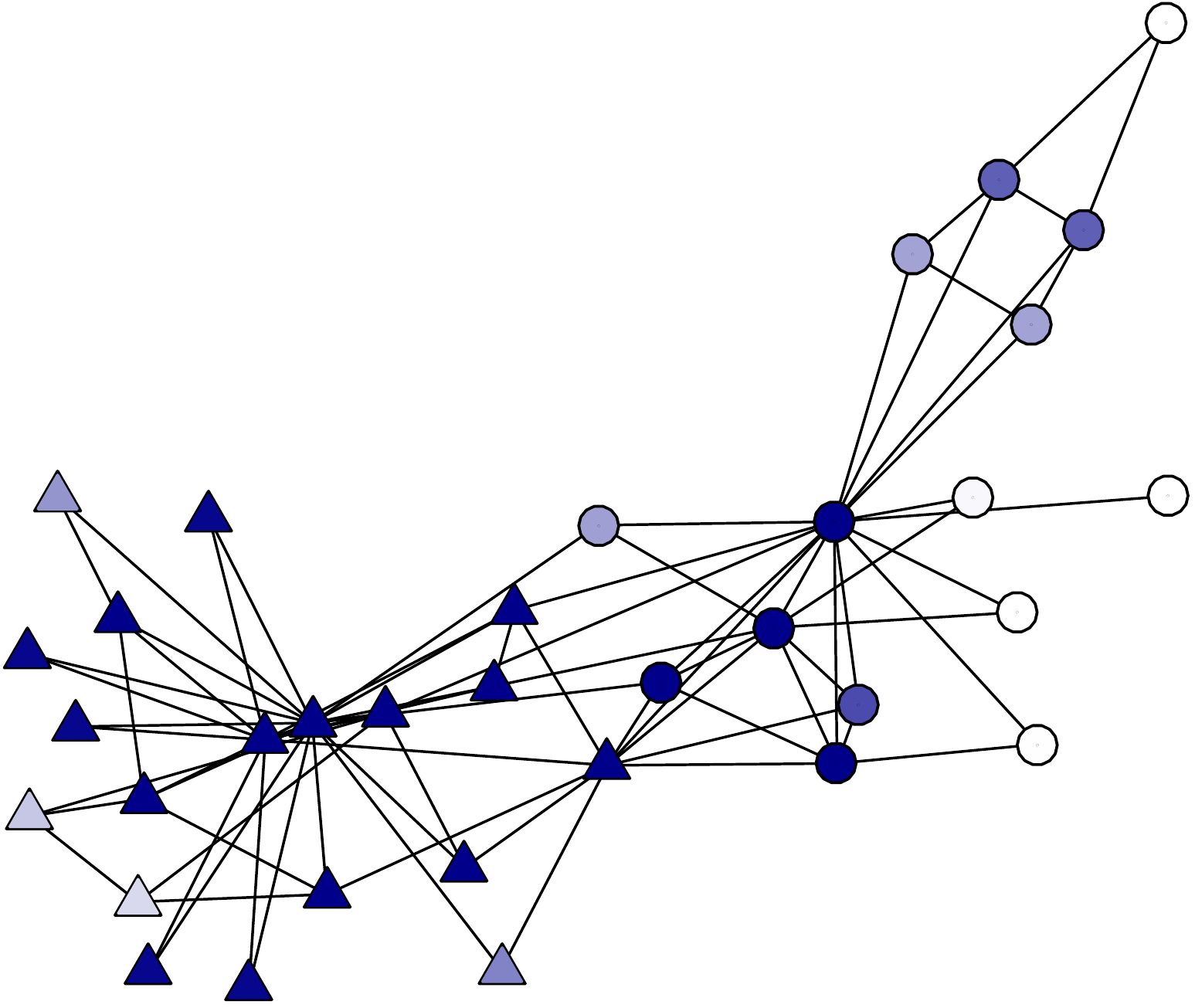}
\caption{Optimal resource allocation for the AMEI Karate Club Network. Left: costs for transmission rates. Right: costs for recovery rates.  Darker colors represent heavier investments, while white nodes do not receive any investment.}
\label{fig:AMEI:resource}
\end{figure}

In this subsection, we illustrate the optimization framework for the optimal
resource allocation over AMEI temporal networks presented in
Theorem~\ref{thm:control:AMEI}. For simplicity in the presentation, and to be
consistent with the Markovian case in the previous section, we focus on the case
where the edge-dynamics $h_{ij}$ of any edge $\{i, j\}$ is a two-state Markov
process taking values in the set $\{1, 2\}$ with the active set $\mathcal A_{ij}
= \{1\}$ and the inactive set $\mathcal D_{ij} = \{2\}$ (although the following
analysis can be applied to the general non-Markovian case where the
edge-dynamics is explained by multi-state Markov processes~$h_{ij}$). In this
subsection, we consider the HeNeSIS spreading model over an AMEI temporal
network based on the static Karate Network. Recall that, in the Markovian Karate
Network, the activations and deactivations of edges within a cluster (or between
clusters) must occur simultaneously. In this numerical simulation, we assume
that these activations and deactivations occur independently of other edges. We
specifically construct our AMEI temporal network as follows. For an edge~$\{i,
j\}$ between the nodes belonging to the first (or second) cluster, we let
$h_{ij}$ be the two-state Markov process whose activation and deactivation rates
are given by~$p_1$ and~$q_1$ ($p_2$ and $q_2$, respectively). Also, we let the
activation and deactivation rates of edges between different clusters to be
$p_3$ and $q_3$. Finally, for a pair $\{i, j\}$ of nodes not connected in the
static Karate Club Network, we let their activation and deactivation rates to be
$0$ and $1$, respectively. This choice guarantees that edges not present in
the static network do not appear in our AMEI Karate network.

Using the cost function in~\eqref{eq:costFunctions}, as well as the box constraints and
budget $\bar R = n/2$ used in Section~\ref{sec:simulation:Markov},
we sub-optimally solve Problem~\ref{prb:epidemiology:AMEI} by using
Theorem~\ref{thm:control:AMEI}. The obtained resource distribution is
illustrated in Fig.~\ref{fig:AMEI:resource}. As for the infection rates, we
observe that nodes at the `boundaries' of the clusters receive small investments, as already observed in the Markovian case. On the other hand, we
cannot clearly observe this phenomena for recovery rates. We finally notice that
the resulting investment heavily leans towards increasing recovery rates, not
decreasing infection rates.

\section{Adaptive Networks} \label{sec:ASIS}

In the case of epidemic outbreaks, it is commonly observed that the connectivity
of human networks is severely influenced by the progress of the disease spread.
This phenomenon, called \emph{social distancing}~\cite{Bell2006,Funk2010}, is
known to help societies cope with epidemic outbreaks. A key
feature of temporal networks of this type is their dependence on the nodal
infection states. However, this structural dependence (or adaptation) cannot be well captured by the Markovian temporal networks because, in
those networks, the dynamics of the network structure is assumed to be independent of the
nodal states. In this direction, the aim of the current section is to present
the so-called \emph{Adaptive SIS} model~\cite{Guo2013,Ogura2015i}, which is able to
replicate adaptation mechanisms found in realistic networks. We first present a
tight extinction condition of epidemic outbreaks evolving in the ASIS model.
Based on this extinction condition, we then illustrate how one can tune the
adaptation rates of networks to eradicate epidemic outbreaks over the ASIS model.

\subsection{Model}

\begin{figure}[tb]
\centering
\includegraphics[width=.65\linewidth]{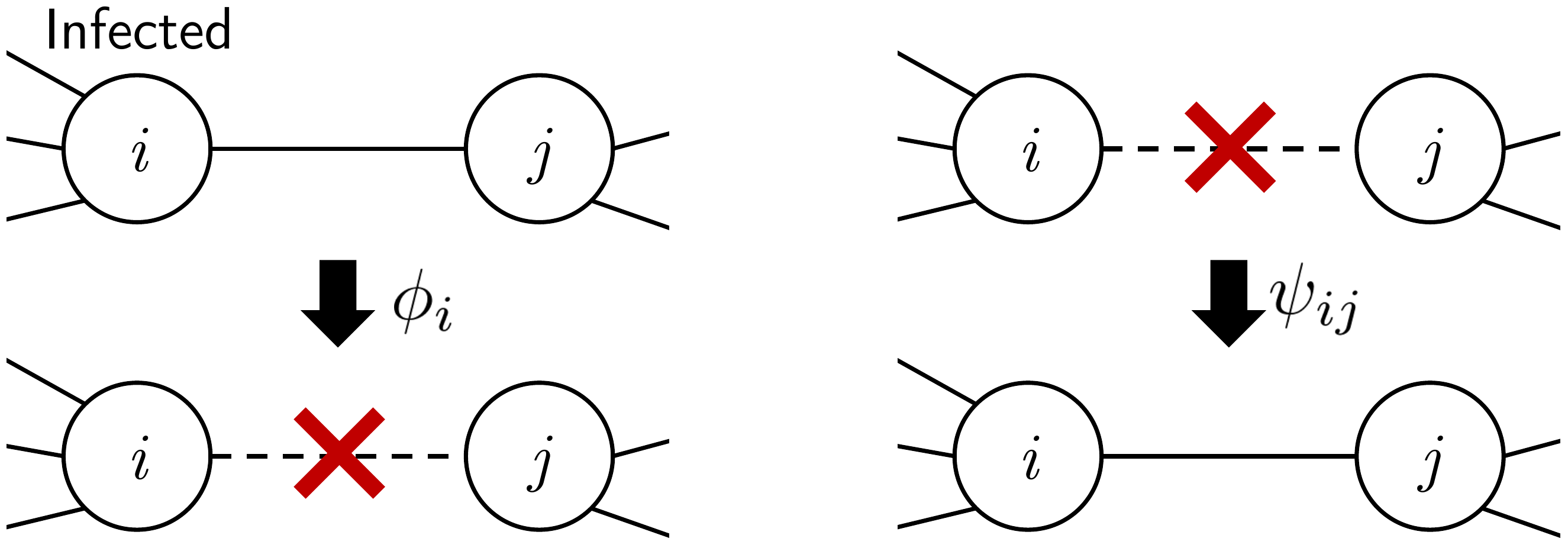}
\caption{Adaptation mechanisms in the Adaptive SIS model.}
\label{fig:adaptive}
\end{figure}

In this section, we first describe the heterogeneous Adaptive SIS (ASIS) model~\cite{Ogura2015i}. As in the HeNeSIS model over Markovian temporal networks (studied
in the previous two sections), the ASIS model consists of the following two
components: the $\{0, 1\}$\nobreakdash-valued infectious, nodal states $x_i(t)$
and a temporal network $\mathcal G(t)$. While the nodal states in the ASIS model
have the same transition probabilities as in \eqref{eq:TVSIS}, the transition
probabilities of the network $\mathcal G(t)$ in the ASIS model are quite
different from Markovian temporal networks because the probabilities depend on
the states of the nodes, as described below. Let $\mathcal G(0) =
(\mathcal V, \mathcal E(0))$ be an initial connected contact graph with
adjacency matrix~$A(0) = [a_{ij}(0)]_{i,j}$. Then, edges in the initial
graph~$\mathcal G(0)$ appear and disappear over time according to the following transition
probabilities:
\begin{align}
\Pr(a_{ij}(t+h) = 0 \mid a_{ij}(t) = 1)
&=
\phi_ix_i(t) h+ \phi_jx_j(t) h +  o(h),
\label{eq:cut}
\\
\Pr(a_{ij}(t+h) = 1 \mid a_{ij}(t) = 0) 
&= 
a_{ij}(0)\psi_{ij} h +  o(h),
\label{eq:rewire}
\end{align}
where the parameters $\phi_i > 0$ and $\psi_{ij} = \psi_{ji}> 0$ are called the
\emph{cutting} and \emph{reconnecting} rates, respectively. Notice that the transition rate in
\eqref{eq:cut} depends on the nodal states~$x_i$ and~$x_j$, inducing an
adaptation mechanism of the network structure to the state of the epidemics. The
transition probability in~\eqref{eq:cut} can be interpreted as a protection
mechanism in which edge $\{i,j\}$ is stochastically removed from the network if
either node~$i$ or $j$ is infected. More specifically, because of the first
summand (respectively, the second summand) in~\eqref{eq:cut}, whenever node~$i$
(respectively, node~$j$) is infected, edge~$\{i, j\}$ is removed from the
network according to a Poisson process with rate~$\phi_i$ (respectively,
rate~$\phi_j$). On the other hand, the transition probability in
\eqref{eq:rewire} describes a mechanism for which a `cut' edge $\{i, j\}$ is
`reconnected' into the network according to a Poisson process with rate
$\psi_{ij}$ (see Fig.~\ref{fig:adaptive}). Notice that we include the
term~$a_{ij}(0)$ in~\eqref{eq:rewire} to guarantee that only edges present in
the initial contact graph~$\mathcal G(0)$ can be added later on by the
reconnecting process. In other words, we constrain the set of edges in the
adaptive network to be a part of the arbitrary contact graph~$\mathcal G(0)$.

\subsection{Optimal Resource Allocation} 

In this section, we consider the situation in which we can tune the values of the cutting rates in
the network by incurring a cost. In particular, we can tune the
value of the cutting rate of node~$i$ to $\phi_i$ by incurring a cost of~$h(\phi_i)$. The total tuning cost is therefore given by $R = \sum_{i=1}^n h(\phi_i)$. In this setup, we can state the optimal resource allocation problem, as follows:

\begin{problem}\label{prb:epidemiology:ASIS}
Consider a heterogeneous ASIS model. Given a budget $\bar R$, tune the cutting
rates $\phi_i$ in the network in such a way that the exponential decay rate of
the infection probabilities is minimized while satisfying the budget constraint
$R\leq \bar R$ and the box-constraint~$0<\ubar \phi \leq \phi_i \leq
\bar{\phi}$.
\end{problem}

In order to solve this problem, we shall follow the same path as we did in the
previous sections: we first find an analytical estimate of the decay rate of the
infection probabilities in the ASIS model. For this purpose, we first represent
the ASIS model by a set of stochastic differential equations described below
(see~\cite{Ogura2015i} for details). For $\gamma > 0$, let $N_\gamma$ denote a
Poisson counter with rate~$\gamma$~\cite{Feller1956vol1}. Then, from the two
equations in~\eqref{eq:TVSIS}, the evolution of the nodal states can be exactly
described by the following stochastic differential equation:
\begin{equation}\label{eq:dx_i}
dx_i 
= 
-x_i \dN_{\delta_i} + (1-x_i) \sum_{j\in\mathcal N_i(0)}  a_{ij}x_j
\dN_{\beta_i},
\end{equation}
for all nodes $i$. Similarly, from \eqref{eq:cut} and \eqref{eq:rewire}, the
evolution of the edges can be exactly described by
\begin{equation}\label{eq:da_ij}
da_{ij} = -a_{ij} (x_i \dN_{\phi_i} + x_j \dN_{\phi_j} )+ (1-a_{ij}) \dN_{\psi_{ij}},
\end{equation}
for all $\{i, j\}\in \mathcal E(0)$.
Then, by \eqref{eq:dx_i}, the expectation $E[x_i]$ obeys the differential
equation $(d/dt)E[x_i] = -\delta_i E[x_i] + \beta_i \sum_{{k\in\mathcal N_i(0)}}
E[(1-x_i)a_{ik}x_k]$. Let $p_i(t) = E[x_i(t)]$ and $q_{ij}(t) =
E[a_{ij}(t)x_i(t)]$. Then, it follows that
\begin{equation}\label{eq:dpidt}
\frac{dp_i}{dt}
= 
-\delta_i p_i + \beta_i \sum_{{j\in\mathcal N_i(0)}} q_{ji}
-
f_i,
\end{equation}
where $f_i = \beta_i \sum_{k\in\mathcal N_i(0)} E[x_ix_ka_{ik}]$ contains
positive higher-order terms. Similarly, from \eqref{eq:dx_i} and
\eqref{eq:da_ij}, the Ito formula for stochastic differential equations (see,
e.g.,~\cite{Hanson2007}) shows that
\begin{align}\label{eq:qijdt}
\frac{dq_{ij}}{dt} = & -\phi_i p_{ij} +  \psi_{ij} (p_i - q_{ij}) 
 - \delta_i q_{ij} + \beta_i \ \sum_{{k\in\mathcal N_i(0)}}\ q_{ki}
- g_{ij},
\end{align}
where $g_{ij} = \phi_j E[x_ix_ja_{ij}] +\beta_i \sum_{{k\in\mathcal N_i(0)}} E[
x_ix_ka_{ik} + (1-a_{ij})a_{ik}x_k]$ contains positive higher-order terms. We
remark that the differential equations~\eqref{eq:dpidt} and \eqref{eq:qijdt}
\emph{exactly} describe the joint evolution of the spreading process and the
network structure without relying on mean-field approximations.

Based on the above derivation, we are able to prove the following proposition: 

\begin{proposition}[\cite{Ogura2015i}]\label{prop:analysis:ASIS}
Let $T_i$ be the unique row-vector satisfying $T_i q = \sum_{{k \in \mathcal
N_i(0)}} q_{ki}$. Define the matrices
\begin{equation*}
\begin{gathered}
B_1 = \col_{1\leq i\leq n} (\beta_i T_i),\quad
B_2 = \col_{1\leq i\leq n} (\beta_i \onev_{d_i} \otimes  T_i),\quad
D_1 = \bigoplus_{i=1}^n\delta_i,\quad 
D_2 = \bigoplus_{i=1}^n( \delta_i I_{d_i}), 
\\
\Phi = \bigoplus_{i=1}^n (\phi_i I_{d_j}),\quad
\Psi_1 = \bigoplus_{i=1}^n (\col_{j\in{\mathcal N}_i(0)}\psi_{ij}),\quad
\Psi_2 = \bigoplus_{i=1}^n \bigoplus_{j\in\mathcal N_i(0)}\psi_{ij},
\end{gathered}
\end{equation*}
where $d_i$ denotes the degree of node~$i$ in the initial graph~$\mathcal G(0)$,
$\otimes$ denotes the Kronecker product~\cite{Brewer1978} of matrices, and
$\onev_n$ denotes the all-one vector of length $n$. If the matrix 
\begin{equation*}
\mathcal A_3  = 
\begin{bmatrix}
-D_1	&B_1
\\
\Psi_1	&B_2 - D_2 - \Phi - \Psi_2
\end{bmatrix}
\end{equation*}
satisfies
\begin{equation}\label{eq:thr}
\lambda_{\max}(\mathcal A_3) < 0, 
\end{equation}
then the infection probabilities in the heterogeneous ASIS model converge to zero exponentially fast with an exponential decay rate~$|\lambda_{\max}(\mathcal A_3)|$.
\end{proposition}

We remark that, in the homogeneous case, where all the nodes share the same
infection rate~$\beta > 0$ and recovery rate~$\delta>0$, and all the edges share
the same cutting rate~$\phi > 0$ and reconnecting rate~$\psi > 0$, the
condition in~\eqref{eq:thr} reduces~\cite{Ogura2015i} to the following simple
inequality:
\begin{equation}\label{eq:threshold:homo}
\frac{\beta}{\delta} < \frac{1 + \omega}{\lambda_{\max}(A(0))}, 
\end{equation}
where $\omega = {\phi}/{(\delta+\psi)}$ is called the \emph{effective cutting
rate}. The proof of this reduction can be found in
\cite[Appendix~B]{Ogura2015i}. We remark that, in the special case when the
network does not adapt to the prevalence of infection, i.e., when $\phi=0$, we
have that $\omega = 0$ and, therefore, the condition
in~\eqref{eq:threshold:homo} is identical to the extinction
condition~\eqref{LinearMFA:TI} corresponding to the homogeneous networked SIS
model over a static network~\cite{VanMieghem2009a}.

Now, based on Proposition~\ref{prop:analysis:ASIS}, one can yield the following
solution to Problem~\ref{prb:epidemiology:ASIS} based on geometric programs:

\begin{theorem}[{\cite[Section~IV]{Ogura2015i}}]\label{thm:control:ASIS}
Assume that there exists $\hat \phi > \bar \phi$ such that the function $\tilde
h(\tilde \phi) = h(\hat \phi - \tilde \phi)$ is a posynomial in $\tilde \phi$.
Then, the cutting rates that sub-optimally solve
Problem~\ref{prb:epidemiology:ASIS} are given by $\{\hat \phi - \tilde
\phi_i^\star\}_{i=1}^n$, where the starred variables solve the optimization
problem:
\begin{equation*}
\begin{aligned}
\minimize_{\tilde \phi_i,\,v>0,\,\lambda>0}\ \ \ 
& 
1/\lambda
\\
\text{\upshape subject to}\ \ \ 
&\mathcal A_3v \leq -\lambda v, 
\\
& \sum_{i=1}^n \tilde h(\tilde \phi_i) \leq \bar R, 
\\
&
\hat{\phi} - \bar{\phi} \leq \tilde{\phi}_i \leq \hat{\phi} - \ubar{\phi}.
\end{aligned}
\end{equation*}
Moreover, this optimization problem can be equivalently converted to a geometric
program.
\end{theorem}

\subsection{Numerical simulations} 

\begin{figure}[tb]
\centering
\includegraphics[width=.75\linewidth]{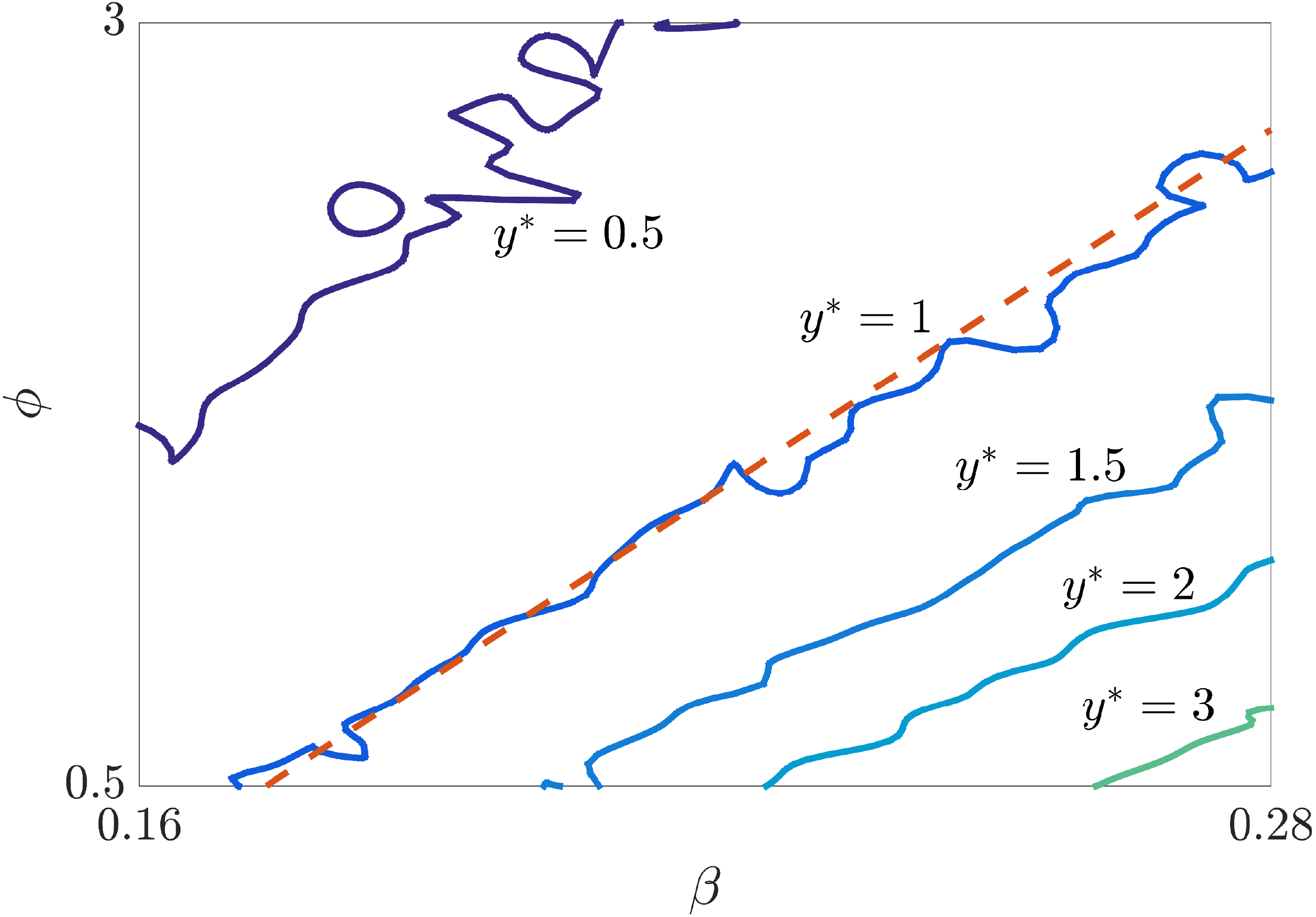}
\caption{The meta-stable number of the infected nodes $y^*$ versus $\beta$ and $\phi$, with $\delta = 1$ and $\psi = 2$. The dashed straight lines show the analytically derived lower bound $(1+\omega)/\lambda_{\max}(A(0)) = \beta$ on the epidemic threshold.}
\label{fig:analysis:ASIS}
\end{figure}

In this section, we illustrate the results presented in this section. Let the
initial graph~$\mathcal G(0)$ be the Karate Club Network. We first consider the
homogeneous case, and fix the recovery rate and the reconnecting rate in the
network to be $\delta = 1$ and $\psi=2$ for all nodes in the graph, for the
purpose of illustration. We then compute the meta-stable number $y^*$ of the
infected nodes in the network for various values of~$\beta$ and $\phi$ (for
details of this simulation, see~\cite{Ogura2015i}). The obtained metastable
numbers are shown as a contour plot in Fig.~\ref{fig:analysis:ASIS}. We see how
the analytical threshold $\beta/\delta = (1+\omega)/\lambda_{\max}(A(0))$ from
\eqref{eq:threshold:homo} (represented as a dashed straight line in
Fig.~\ref{fig:analysis:ASIS}) is in good accordance with the numerically found
threshold $y^* = 1$.

\begin{figure}[tb]
\centering
\includegraphics[width=\resourceGraphWidth]{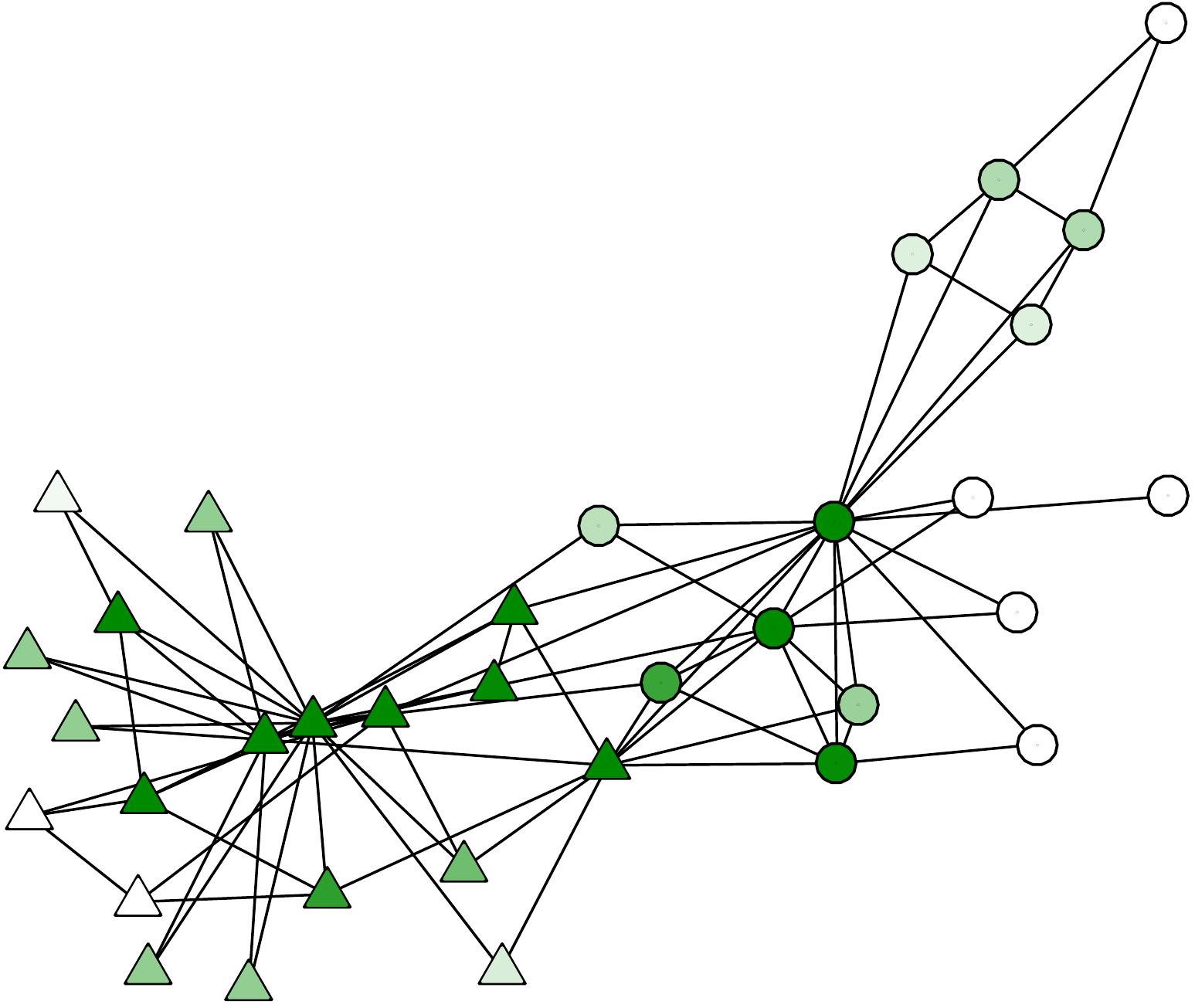}
\caption{The distribution of the resource for tuning the cutting rates in the ASIS Karate Network}
\label{fig:ASIS:allocation}
\end{figure}

We then consider the optimal resource distribution problem stated in
Problem~\ref{prb:epidemiology:ASIS}. In this simulation, we use the cost
function $h(\phi) = c_5+c_6/(\hat \phi - \phi)^s$ similar to the one used
in~\eqref{eq:costFunctions}, where $s$ is a positive parameter for tuning the
shape of the cost function, $\hat \phi$ is a constant larger than $\bar \phi$,
and $c_5$ and $c_6$ are constants such that $h(\ubar \phi) = 0$ and $h(\bar
\phi) = 1$. We let $\ubar\phi = 0.5$, $\bar \phi = 1.5$, $\hat\phi = 100\bar
\phi$, and $s = 1$, for which the resulting cost function resembles a linear
function as in the case of Markovian temporal networks. Using these cost
functions and the budget $\bar R = n/2$, we solve the optimization problem in
Theorem~\ref{thm:control:ASIS} to find the sub-optimal distribution of resource
over the network (illustrated in Fig.~\ref{fig:ASIS:allocation}). Interestingly,
unlike in the Markovian cases in Sections~\ref{sec:Markov} and~\ref{sec:AMEI},
we cannot clearly observe the phenomenon where nodes at the boundaries of the
clusters receive relatively less investments.

\begin{figure}[tb]
\centering
\includegraphics[width=.75\textwidth]{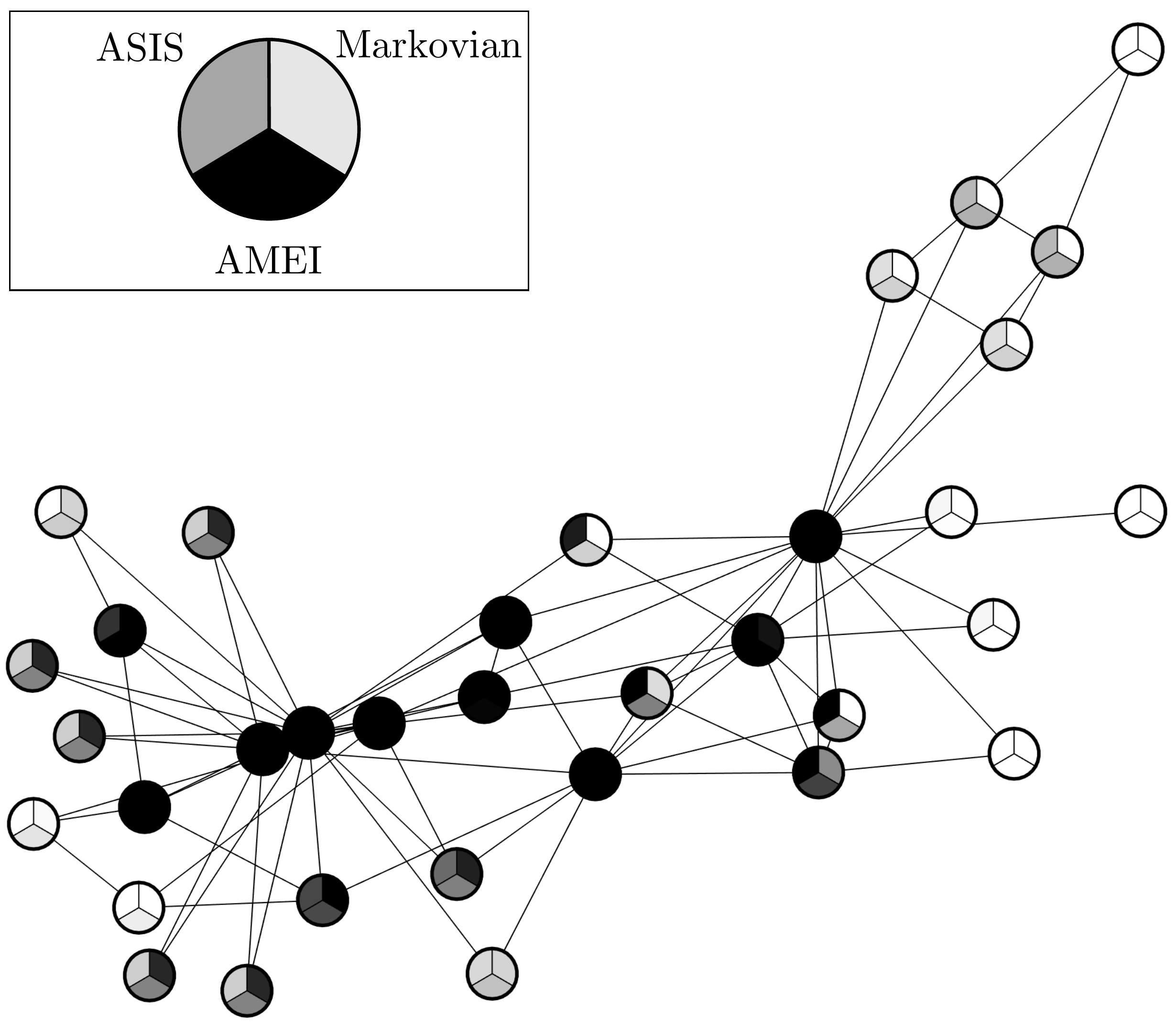}
\caption{Optimal spending on nodes. In each circle (node), the upper-right, lower, and upper-left colors indicate the investments according to the Markovian, AMEI, and ASIS formulations, respectively.}
\label{fig:allallocations}
\end{figure}

Finally, in Fig.~\ref{fig:allallocations}, we summarize the amount of optimal
resource distributions obtained for the Markovian, AMEI, and ASIS Karate
Networks. We see that, although the three allocations share a certain tendency
such as concentration of resource on high-degree nodes, they are not necessarily
qualitatively equal. This observation confirms the necessity of appropriately
incorporating the characteristics of temporal/adaptive networks into our
mechanism of resource distributions.

\section{Conclusion}

In this chapter, we have given an overview of recent progress on the problem of containing  epidemic outbreaks taking place in temporal and adaptive
complex networks. Specifically, we have presented analytical frameworks for finding the optimal distribution of resources over Markovian temporal networks, aggregated-Markovian
edge-independent temporal networks, and in the Adaptive SIS model. For each of
the cases, we have seen that the optimal resource distribution problems can be
reduced to an efficiently solvable class of convex optimizations called
geometric programs. We have illustrated the results with several numerical
simulations based on the well-studied Zachary Karate Club Network.

\begin{acknowledgement}
This work was supported in part by the NSF under Grants
CNS-1302222 and IIS-1447470.
\end{acknowledgement}
%


\end{document}